\documentclass[%
 reprint,
 amsmath,amssymb,
 aps,
]{revtex4-2}

\usepackage[english]{babel}
\usepackage{graphicx,import}
\usepackage{dcolumn}
\usepackage{bm}

\usepackage[utf8]{inputenc}
\usepackage[T1]{fontenc}
\usepackage{mathptmx}
\usepackage{etoolbox}

\usepackage{tikz}
\usetikzlibrary{shadows}
\usetikzlibrary{matrix}
\usetikzlibrary{fit}
\usetikzlibrary{backgrounds}
\usetikzlibrary{positioning}
\usetikzlibrary{arrows}
\usetikzlibrary{shapes}
\usetikzlibrary{calc}
\usepackage{circuitikz}
\usepackage{setspace}
\usepackage{multirow}
\usepackage{subcaption}
\usepackage{booktabs}
\usepackage{adjustbox}
\usetikzlibrary{decorations.pathreplacing, angles, quotes}

\usetikzlibrary{arrows.meta, positioning, chains}
\usepackage{tikz}
\usetikzlibrary{matrix}
\usetikzlibrary{fit}
\usetikzlibrary{backgrounds}
\usetikzlibrary{positioning}
\usetikzlibrary{arrows}
\usetikzlibrary{shapes}
\usetikzlibrary{calc}
\usepackage{circuitikz}
\usepackage{setspace}

\tikzstyle{state}=[rectangle,
    thick,
    minimum height=0.6cm,
    minimum width=1.0cm,
    rounded corners]

\tikzstyle{state1}=[state,
    draw=orange!80,
    fill=orange!20]

\tikzstyle{state2}=[state,
    draw=teal!80,
    fill=teal!20]

\tikzstyle{state3}=[state,
    draw=blue!80,
    fill=blue!20]

\tikzstyle{state4}=[state,
    draw=brown!80,
    fill=brown!20]

\tikzstyle{state5}=[state,
    draw=yellow!80,
    fill=yellow!20]

\tikzstyle{line} = [draw, -latex']

\tikzstyle{auto}=[rectangle,
    thick,
    minimum height=0.6cm,
    minimum width=0.6cm,
    draw=purple!80,
    fill=purple!20,
    rounded corners]

\tikzstyle{background}=[rectangle,
    fill=gray!10,
    inner sep=0.15cm,
    rounded corners=3mm]

\begin{document}

\preprint{APS/123-QED}

\title{Attention-Enhanced Reservoir Computing as a Multiple Dynamical System Approximator}

\author{Felix Köster}
 \email{felixk@mail.saitama-u.ac.jp}
\author{Kazutaka Kanno}%
\author{Atsushi Uchida}%
 \email{auchida@mail.saitama-u.ac.jp}
\affiliation{%
 Department of Information and Computer Sciences, Saitama University 255 Shimo-Okubo, Sakura-ku, Saitama City, Saitama, 338–8570, Japan
}%

\date{\today}

\begin{abstract}
Reservoir computing has proven effective for tasks such as time-series prediction, particularly in the context of chaotic systems. However, conventional reservoir computing frameworks often face challenges in achieving high prediction accuracy and adapting to diverse dynamical problems due to their reliance on fixed weight structures. A concept of an attention-enhanced reservoir computer has been proposed, which integrates an attention mechanism into the output layer of the reservoir computing model. This addition enables the system to prioritize distinct features dynamically, enhancing adaptability and prediction performance. In this study, we demonstrate the capability of the attention-enhanced reservoir computer to learn and predict multiple chaotic attractors simultaneously with a single set of weights, thus enabling transitions between attractors without explicit retraining. The method is validated using benchmark tasks, including the Lorenz system, Rössler system, Henon map, Duffing oscillator, and Mackey-Glass delay-differential equation. Our results indicate that the attention-enhanced reservoir computer achieves superior prediction accuracy, valid prediction times, and improved representation of spectral and histogram characteristics compared to traditional reservoir computing methods, establishing it as a robust tool for modeling complex dynamical systems.
\end{abstract}

\maketitle

\section{\label{Sec1}Introduction}

The combination of reservoir computing (RC) and other machine learning concepts has emerged as a promising field of information processing, utilizing the development of more rapid and efficient computational schemes \cite{shen2017deep}. Central to this area is the concept of RC, a scheme that utilizes the dynamic properties of any dynamical system \cite{jaeger2001echo, appeltant2011information, tanaka2019recent, torrejon2017neuromorphic, kotooka2021}. RCs leverage a fixed, randomly connected reservoir and train only the output layer, making them computationally efficient and suitable for a variety of tasks, including time-series prediction and pattern recognition \cite{jaeger2001echo, tanaka2019recent}, while making hardware implementations easy and versatile.

Chaotic time-series prediction is known to be challenging owing to its inherent sensitivity to small deviations \cite{strogatz:2000}. A prime example of this challenge is weather forecasting, which is typically limited to short-term predictions due to the fundamentally chaotic nature of weather systems.

The framework of RC has been widely implemented using a variety of physical devices, such as electronic circuits \cite{appeltant2011information}, spintronic devices \cite{torrejon2017neuromorphic}, soft robots \cite{ef0a6b769141456283377bec586ba791}, and nanostructure materials \cite{kotooka2021}, showing the broad application range. It also highlights that many reservoir computer can be leveraged in a wide area, where the utilization can often easily be accomplished via a simple measurement process.
One of the advantages of RC is the easy implementation of the reservoir without high computational cost because input weights and network weights are randomly fixed, and only the output weights are trained using easy and fast ridge regression, thus, making the output layer linear. Therefore, classical RC is often limited due to inflexibility of the fixed output weight structure when faced with the complexities in chaotic time series \cite{jaeger2001echo}, while also having a bad scalability due to the characteristics of ridge regression and linear output layers. 

Advancements in neural network architectures, such as the attention mechanism, have opened new pathways for enhanced prediction accuracy and improved adaptability \cite{vaswani2017attention, bai2018empirical}.
To further enhance the predictive capabilities of RC, we proposed integrating the attention mechanism within a photonic setup \cite{koester2024}. By incorporating an attention layer as the output layer, the system gets the capacity to prioritize specific readouts in each time step, potentially yielding a more dynamic comprehension of the data \cite{bai2018empirical}. 

In this study, we use an attention-enhanced reservoir computer (AERC) with an echo state network to demonstrate its capability to learn multiple attractors simultaneously with the same set of weights and to acquire the emergent ability to switch between learned attractors without explicit training. We perform chaotic time-series prediction by deploying five different benchmark tasks, i.e. prediction of the Lorenz system \cite{lorenz1963deterministic}, Rössler system \cite{rossler}, Henon map \cite{henon}, Duffing oscillator \cite{duffing}, and Mackey-Glass delay differential equation \cite{mackey}, showcasing the robustness and adaptability of the attention mechanism. These diverse tasks allow us to investigate how well the attention mechanism adapts to various types of chaotic dynamics and how effectively it improves prediction accuracy. By training the AERC on these five different tasks with a single set of weights, we demonstrate its ability to simultaneously learn and predict multiple attractors, a significant advancement over traditional RC.

\section{Reservoir Computing and Attention Mechanism}

\subsection{Classic Reservoir Computing}

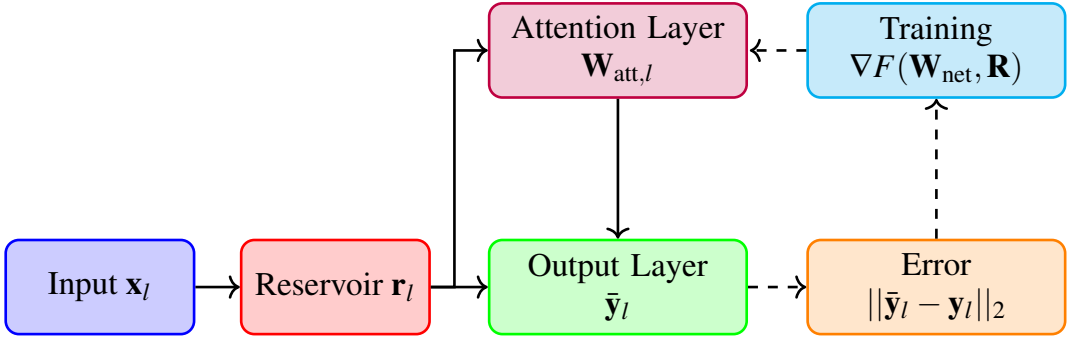
\begin{figure*}
\centering
\resizebox{0.8\textwidth}{!}{
    \begin{tikzpicture}[node distance=2cm, thick]

    \tikzstyle{input_style} = [draw=blue, fill=blue!20, rounded corners, rectangle, minimum width=2cm, minimum height=1cm]
    \tikzstyle{reservoir_style} = [draw=red, fill=red!20, rounded corners, rectangle, minimum width=2cm, minimum height=1cm]
    \tikzstyle{output_style} = [draw=green, fill=green!20, rounded corners, rectangle, minimum width=2cm, minimum height=1cm]
    \tikzstyle{attention_style} = [draw=purple, fill=purple!20, rounded corners, rectangle, minimum width=2cm, minimum height=1cm]
    \tikzstyle{prediction_style} = [draw=orange, fill=orange!20, rounded corners, rectangle, minimum width=2cm, minimum height=1cm]
    \tikzstyle{training_style} = [draw=cyan, fill=cyan!20, rounded corners, rectangle, minimum width=2cm, minimum height=1cm]

    \node (input) [input_style] {Input $\mathbf{x}_l$};
    \node (reservoir) [reservoir_style, right of=input, xshift=0.5cm] {Reservoir $\mathbf{r}_l$};
    \node (output) [output_style, right of=reservoir, xshift=1cm, anchor=center, text width=2.5cm, align=center] {Output Layer \\ $\bar{\mathbf{y}}_l$};
    \node (attention) [attention_style, above=2cm of output, anchor=center, text width=2.5cm, align=center] {Attention Layer \\ $\mathbf{W}_{\text{att},l}$};
    \node (prediction) [prediction_style, right=2cm of output, anchor=center, text width=2.5cm,, align=center] {Error  \\ $||\bar{\mathbf{y}}_l - \mathbf{y}_l||_2$};
    \node (training) [training_style, right=2cm of attention, anchor=center, text width=2.5cm, align=center] {Training \\ $\nabla F(\textbf{W}_{\text{net}}, \textbf{R})$};

    \draw[->] (input) -- (reservoir) node[midway, above] {}; 
    
    \draw[->] (reservoir.east) -- ++(0.25,0) coordinate (mid) |- (attention.west);

    \draw[->] (attention.south) -- (output.north) node[midway, right] {};
    \draw[->] (reservoir) -- (output) node[midway, above] {};
    \draw[->, dashed] (output) -- (prediction) node[midway, above] {};
    \draw[->, dashed] (prediction.north) -- (training.south) node[midway, right] {};
    
    \draw[->, dashed] (training.west) -- ++(-0.6,0);  
    
\end{tikzpicture}
}
\caption{The float chart diagram of the AERC layout. Via the inputs reservoir states are collected. With the reservoir states and a randomly initialized attention layer an output is generated, which is then utilized in a gradient descent approach to optimize the attention layer. The dashed arrows show the training process, which are deactivated after training.}
\label{fig:float_diagram}
\end{figure*}

RC is a framework for computation that utilizes a fixed, randomly connected, and high-dimensional dynamical system, called a reservoir, along with a trained readout layer to perform tasks such as time series prediction, pattern recognition, and classification.

Mathematically, the state of the reservoir at time-step \( l \), denoted by \( \mathbf{r}_l \), is updated based on the previous state and the current input \( \mathbf{x}_l \) as follows:
\[
\mathbf{r}_l = \tanh(\mathbf{W}_{\text{res}} \mathbf{r}_{l-1} + \mathbf{W}_{\text{in}} \mathbf{x}_l + \mathbf{b})
\]
Here, \( \mathbf{r}_l \) is the reservoir state vector in \( \mathbb{R}^{N\times1} \), where \( N \) represents the number of reservoir nodes or readouts. The matrix \( \mathbf{W}_{\text{res}} \) in \( \mathbb{R}^{N \times N} \) is the internal reservoir weight matrix, \( \mathbf{W}_{\text{in}} \) in \( \mathbb{R}^{N \times M} \) is the input weight matrix, and \( \mathbf{x}_l \) in \( \mathbb{R}^{M\times1} \) is the input vector, with \( M \) representing the number of input dimensions. The vector \( \mathbf{b} \) in \( \mathbb{R}^{N\times1} \) is the bias vector.

The linear readout layer is trained to map the reservoir states to the desired output. The output \( \bar{\mathbf{y}}_l \) at the \( l \)-th data point is given by:
\[
\bar{\mathbf{y}}_l = \mathbf{W}_{\text{out}} \mathbf{r}_l
\]
where \( \mathbf{W}_{\text{out}} \) in \( \mathbb{R}^{T \times N} \) is the output weight matrix, with \( T \) representing the number of output dimensions.

The feature matrix \( \mathbf{R} \), used for training the readout layer, consists of reservoir states collected over time. Each row of \( \mathbf{R} \) represents a sample (a reservoir state at a particular time-step), and each column represents a feature (a particular node of the reservoir state vector):
\[
\mathbf{R} = \begin{bmatrix}
\mathbf{r}_1^\top \\
\mathbf{r}_2^\top \\
\vdots \\
\mathbf{r}_N^\top
\end{bmatrix}
= \begin{bmatrix}
r_{1,1} & r_{1,2} & \cdots & r_{1,L} \\
r_{2,1} & r_{2,2} & \cdots & r_{2,L} \\
\vdots & \vdots & \ddots & \vdots \\
r_{N,1} & r_{N,2} & \cdots & r_{N,L}
\end{bmatrix}
\]
where \( L \) is the total number of measured time steps, and \( N \) is the number of reservoir nodes.

The readout weights \( \mathbf{W}_{\text{out}} \) are typically trained using linear regression to minimize the error between the predicted output \( \bar{\mathbf{y}}_l \) and the desired output \( \mathbf{y}_l \):
\[
\mathbf{W}_{\text{out}} =  (\mathbf{R}^\top \mathbf{R} + \lambda \mathbf{I})^{-1}\mathbf{R}^\top\mathbf{Y} 
\] 
where \( \mathbf{Y} \) is the matrix of desired outputs, \( \lambda \) is a regularization parameter, and \( \mathbf{I} \) is the identity matrix.

Throughout this paper, we fixed the spectral radius of the reservoirs to $0.9$ and the average connection degree of the input matrix and topology matrix to $0.1$.

\begin{table*}
    \centering
    \begin{tabular}{lcccc}
        \toprule
        \textbf{System} & \textbf{Total Time} & \textbf{Sample Points} & \textbf{Step Size} & \textbf{Samples in one Lyapunov Time}\\
        \midrule
        Lorenz & 375 & 7500 & 0.05 & 22 \\
        Rössler Attractor & 2000 & 7500 & 0.27 & 52 \\
        Henon Map & 7500 & 7500 & 1 & 9.5 \\
        Duffing Oscillator & 825 & 7500 & 0.11 & 21.6\\
        Mackey-Glass & 7500 & 7500 & 1 & 166.7\\
        \bottomrule
    \end{tabular}
    \caption{Sampling step sizes for each dynamical system.}
    \label{tab:sampling_steps}
\end{table*}

\subsection{Attention-Enhanced Reservoir Computing}

Incorporating an attention mechanism into a RC can significantly improve its performance by making the output weights adaptable to the reservoir states, however, also increasing the number of trainable weights. Instead of using fixed output weights as in traditional RC, the attention-enhanced model uses dynamically computed attention weights that vary with the inputs \cite{koester2024}.

For each data point \( l \) fed into the reservoir, the corresponding attention weights \( \mathbf{W}_{\text{att}, l} \) in \( \mathbb{R}^{K\times N} \)  are computed. These attention weights are derived using a neural network \( F \) with trainable parameters \( \mathbf{W}_{\text{net}} \), based on the reservoir states \( \mathbf{r}_l \):
\[
\mathbf{W}_{\text{att},l} = F(\mathbf{W}_{\text{net}}, \mathbf{r}_l)
\]
\( \mathbf{W}_{\text{net}} \) is here a set of matrices that collectively build up the whole neural network.
The output \( \bar{\mathbf{y}}_l \) for each data point is then calculated as a weighted sum:
\[
\bar{\mathbf{y}}_l = \mathbf{W}_{\text{att},l} \mathbf{r}_l
\]
Here, \( \mathbf{W}_{\text{att},l} \in \mathbb{R}^{T \times N} \) are the input-dependent attention weights, and \( \bar{\mathbf{y}}_l \) represents the approximation for the target signal \( \mathbf{y}_l \).

We train the AERC to predict the next step in the time series.
To optimize the attention layer weights during training, gradient descent is applied to minimize the normalized root mean square error between the predictions $\bar{\mathbf{y}}_l $ and targets $\mathbf{y}_l$ over all data points $L$. 
\begin{align}
    \textbf{W}_{\text{net}}(s+1) = \textbf{W}_{\text{net}}(s) -\gamma \nabla F(\textbf{W}_{\text{net}}(s), \textbf{R}),
\end{align}
where $s$ indexes the epochs used for training, while one epoch is done over the whole dataset.
The goal is to refine \( \mathbf{W}_{\text{net}} \) such that the AERC is capable to capture the complex dependencies in the input data.
In this paper, we use a neural network consisting of one hidden layer with a ReLu activation function. \\
After the training phase, the attention mechanism is used in a closed-loop configuration for autonomous prediction. In this setup, the reservoir is first fed with a few steps of one of the benchmark systems, after which the loop is closed. At each time step, the attention layer computes new attention weights based on the reservoir states. These dynamically updated attention weights allow the system to better adapt to the evolving input, significantly improving prediction performance in complex dynamical systems, and making it possible to dynamically adapt to the input.

\subsection{Flowchart of Attention-Enhanced Reservoir Computing}

Figure \ref{fig:float_diagram} represents the flowchart of the core processes involved in our proposed AERC framework, integrating traditional RC with an attention mechanism to enable a single machine learning agent predicting multiple systems.\\
In the input stage, the standardized input data, denoted as \( \mathbf{x}_l \), is passed into the reservoir. This reservoir, operating as a fixed, high-dimensional, nonlinear dynamic system, processes the data and generates a corresponding reservoir state \( \mathbf{r}_l \), capturing the dynamics of the input sequence up to input time step $l$, with \( \mathbf{r}_l \) being a high dimensional embedding.\\
The next stage, the attention layer distinguishes the AERC from traditional reservoir computer approaches. The attention layer receives the reservoir state \( \mathbf{r}_l \) and computes attention weights \( \mathbf{W}_{\text{att}, l} \), assigning varying levels of importance (attention) to different nodes of the reservoir in every time step based on their relevance to the current prediction step.

After that, the output layer processes the reservoir state \( \mathbf{r}_l \) weighted by \( \mathbf{W}_{\text{att}, l} \), resulting in the prediction output \( \bar{\mathbf{y}}_l \). With the introduction of the attention mechanism, the AERC gains the ability to adjust dynamically to evolving temporal patterns, often improving the quality of its predictions.

Finally, the error is computed and passed to the training process. The training optimizes the attention layer weights $\mathbf{W}_{\text{net}}$ using gradient-based methods to minimize prediction errors.
After the error converges to a lowest value, the training is finished and the trained AERC can be used in single step predictions or a closed loop configuration by reinserting its own prediction as input.

To accelerate the training, We first feed the reservoir with all the data points of the input time series, collecting the reservoir states \( \mathbf{r}_l \) with their corresponding target values \( \mathbf{y}_l \), which in our case is the next step in the time series that was fed to the reservoir up to time step $l$. After having all $L$ reservoir states \( \mathbf{r}_l \) and their targets \( \mathbf{y}_l \), we run the gradient descent algorithm to minimize the defined loss (see Sec. \ref{seq:loss}) via Pytorch.

\section{Prediction of Multiple attractors}
We use an AERC to simultaneously predict multiple different attractors at the same time with the same set of weights.
We show that with increasing reservoir size the valid prediction time increases and the power spectrum and histogram distributions of the predicted time series gets more similar with all true systems. We train an AERC to predict the next step for a set of five different systems simultaneously.
After training, we feed the trained AERC with an initial condition of a selected system and then close the loop for autonomous prediction.
The predicted time series is compared with the true one of the chosen system to get the valid prediction time, the power spectrum similarity, and the histogram similarity.

We simulate five well-known dynamical systems as benchmark tasks: the Lorenz system \cite{lorenz1963deterministic}, Rössler system \cite{rossler}, Henon map \cite{henon}, Duffing oscillator \cite{duffing}, and Mackey-Glass delay differential equation \cite{mackey}. Each system exhibits unique characteristics and operates over different time scales, offering a diverse set of challenges for predictive modeling. The detailed equations of the five dynamical models are shown in the Appendix.

These five dynamical systems consist of very different chaotic system tasks: three-dimensional systems of ordinary nonlinear differential equations, a two-dimensional driven system, a two-dimensional map, and a one-dimensional delay differential equation. Learning all five systems with different time scales at the same time is challenging.

For each dynamical system, time series data are sampled at regular intervals determined by dividing the total simulation time by the number of data points. The sampling step sizes for each system are listed in Table \ref{tab:sampling_steps}.

\begin{figure*}[t]
    \centering
    \includegraphics[width=0.67\textwidth]{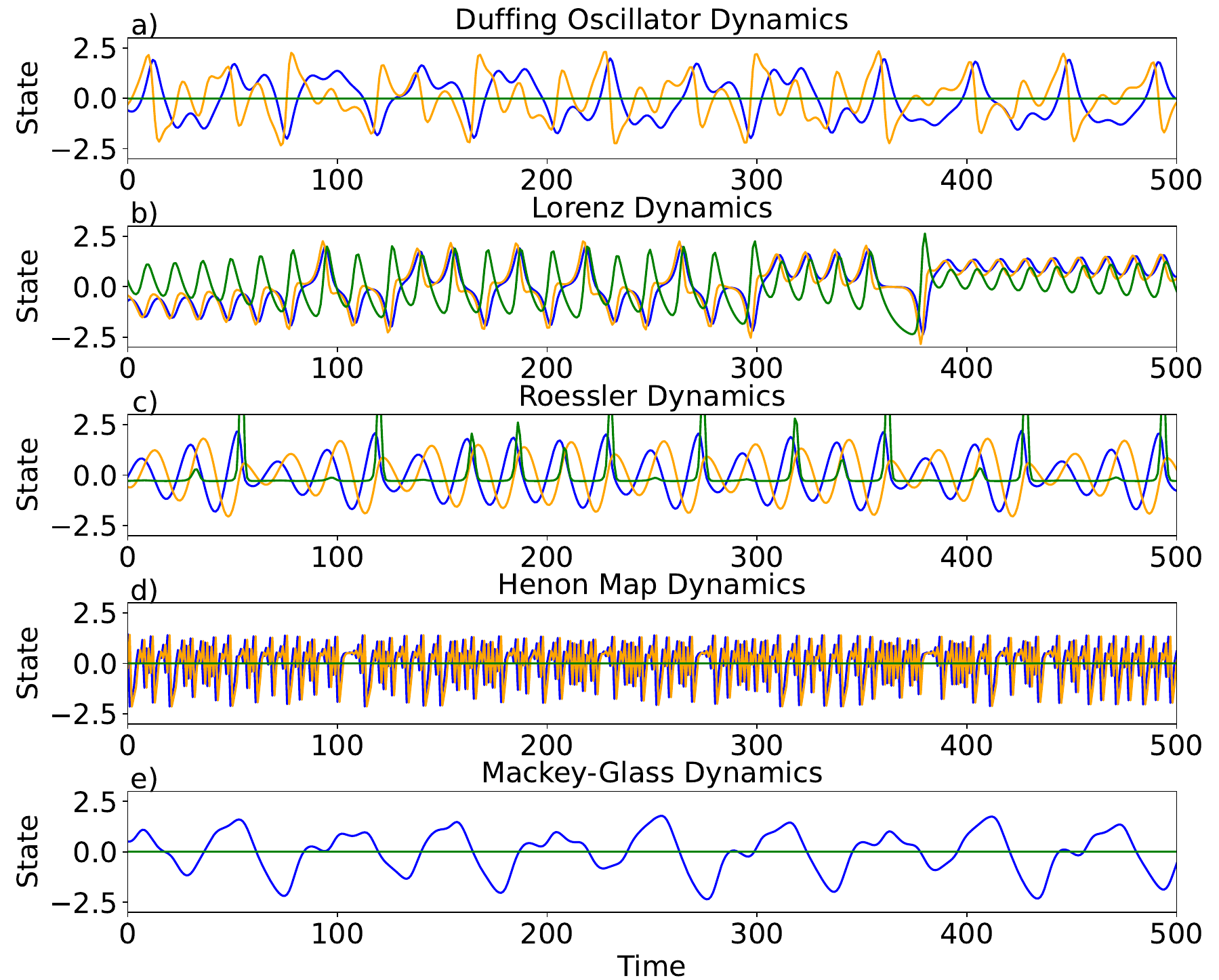}  
    \caption{Example time series from the dataset of all five different tasks after standardization. Each subplot corresponds to one of the dynamical systems simulated: the (a) Duffing oscillator, (b) Lorenz system, (c) Rössler system, (d) Henon map, and (e) Mackey-Glass delay differential equation.}
    \label{fig:time_series_plots}
\end{figure*}

\subsection{Normalization Process}
To ensure that the data from different systems are comparable, we applied standardization to each time series. Standardization transforms the data to have a mean of zero and a standard deviation of one:
\[
\text{x(t)} = \frac{X(t) - \mu}{\sigma}
\]
where \(x(t)\) is the standarized value, \(X(t)\) is the original value, \(\mu\) is the mean, and \(\sigma\) is the standard deviation of the time series.
We make this transformation for all dimensions of all time series individually.

Figure \ref{fig:time_series_plots} shows the time series from the standardized dataset for all five different tasks: the Lorenz system, Rössler system, Henon map, Duffing oscillator, and Mackey-Glass delay differential equation. The standardization process, which transforms the time series to have a mean of zero and a standard deviation of one, allows for the direct comparison of these systems showing the differences in their time scales and state-space amplitudes.

For instance, the Henon map exhibits fast chaotic dynamics, while both the Lorenz and Rössler systems show chaotic and continuous dynamics. On the other hand, the Mackey-Glass system demonstrates slower delayed responses. The variety of patterns present across these systems show the challenge of creating a unified machine learning agent capable of accurately predicting all system dynamics.

\subsection{Dual-Objective Training of Attention-Enhanced Reservoir Computer}
\label{seq:loss}

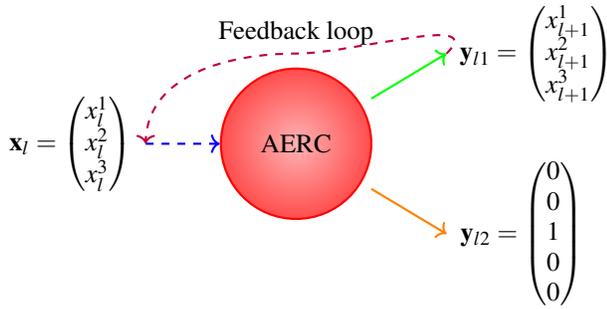
\begin{figure}
    \centering
    \begin{tikzpicture}
        \tikzstyle{input_style} = [draw=blue, fill=blue!20, thick, dashed]
        \tikzstyle{output_style1} = [draw=green, fill=green!20, thick]
        \tikzstyle{output_style2} = [draw=orange, fill=orange!20, thick]
        \tikzstyle{circle_style} = [circle, draw=red, thick, minimum size=2cm, shading=radial, inner color=red!30, outer color=red!70]
        \tikzstyle{feedback_style} = [->, draw=purple, thick, dashed] 

        \node[circle_style] (AERC) at (0,0) {AERC};

        \draw[->, input_style] (-2,0) -- (AERC.west);

        \node at (-3, 0) {$\mathbf{x}_l= \begin{pmatrix}
x^1_l \\
x^2_l \\
x^3_l
\end{pmatrix}$};

        \draw[->, output_style1] (1,0.6) -- (2,1.2); 
        \draw[->, output_style2] (1,-0.6) -- (2,-1.2); 

        \node at (3.2, 1.2) {$\mathbf{y}_{l1}= \begin{pmatrix}
x^1_{l+1} \\
x^2_{l+1} \\
x^3_{l+1}
\end{pmatrix}$};
        \node at (3, -1.2) {$\mathbf{y}_{l2} = \begin{pmatrix}
0\\
0 \\
1 \\
0 \\
0 
\end{pmatrix}$};

        \draw[feedback_style] (2,1.2) to[out=45, in=0] (0,1.2) to[out=180, in=90] (-2,0); 
\node[above] at (0,1.2) {Feedback loop};
    \end{tikzpicture}
    \caption{Diagram of AERC target. The input vector $\mathbf{x}_l$ enters the AERC, and the output consists of two vectors, $\mathbf{y}_{l1}$ and $\mathbf{y}_{l2}$. These vectors represent different aspects of the prediction target. The vector $\mathbf{y}_{l1}$ is used for next-step prediction, while $\mathbf{y}_{l2}$ helps to characterize the attractor that the system is currently on. Together, these two vectors form the complete target.}
    \label{fig:digram_1_AERC_training}
\end{figure}

The AERC is trained to predict the next time step in a dynamical system, and to characterize the attractor that the system is on. In this framework shown in Fig. \ref{fig:digram_1_AERC_training}, the input vector $\mathbf{x}_l$ consisting of the time series enters the AERC, where it is processed to generate two output vectors, $\mathbf{y}_{l1}$ and $\mathbf{y}_{l2}$.

The vector $\mathbf{y}_{l1}$ is responsible for predicting the next time step of the system. On the other hand, the vector $\mathbf{y}_{l2}$ is a one hot vector used to estimate the class of the attractor that the system is on. By simultaneously learning these two target vectors, it is possible to check which attractor the AERC is on.

It is worth noting that the classified output (target) vector $\mathbf{y}_{l2}$ is only used for analyzing the reservoir's performance and is not revealed to the system during training or prediction. This ensures that the AERC cannot "cheat" by knowing which attractor it is on beforehand and must rely solely on the input dynamics to make its predictions.
We will check the AERC performance if the class is also provided as an input in Sec. \ref{sec:dual_input}

To estimate the first vector, $\mathbf{y}_{l1}$, gradient descent is used to train the model by minimizing the Mean Squared Error (MSE) between the predicted and actual signals. For the second vector, $\mathbf{y}_{l2}$, a cross-correlation loss is applied.

By using two simultaneous objectives, we introduce the mean-squared error (MSE) loss, defined as:
\[
L_{\text{MSE}} = \frac{1}{L} \sum_{l=1}^{L} (\mathbf{y}_{l1} - \bar{\mathbf{y}}_{l1})^2
\]
where \( \mathbf{y}_{l1} \) is the true time series, \( \bar{\mathbf{y}}_{l1} \) is the predicted time series, and \( L \) is the total number of time steps.

Simultaneously, the attractor classification is trained using cross-entropy loss as follows.
\[
L_{\text{CE}} = - \frac{1}{L} \sum_{l=1}^{L} \mathbf{y}_{l2} \log(\bar{\mathbf{y}}_{l2})
\]
where \( \mathbf{y}_{l2} \) is the true attractor class probability and \( \bar{\mathbf{y}}_{l2} \) is the predicted class probability.

The total loss is a sum of the two:
\[
L_{\text{total}} = L_{\text{MSE}} + L_{\text{CE}}.
\]

Gradient descent is used to update the network parameters \( \mathbf{W}_{\text{net}} \), minimizing \( L_{\text{total}} \) at each training step. The model is trained over multiple epochs to achieve both accurate next-step predictions and robust attractor classification. After training, the AERC operates autonomously in a closed-loop configuration.

\section{Results of time series prediction}

\subsection{Chaotic Time Series of Different Tasks}

We train an AERC with a neural network consisting of one hidden layer and a ReLU activation function to predict the next time-step for all five different tasks. The closed-loop configuration is employed for autonomous prediction after training the AERC. In the closed-loop setup, the RC is initially fed the first few time steps from one system to initialize the network. After this, the loop is closed, allowing the reservoir to continue predicting the trajectory of the chosen system autonomously without further input.

\begin{figure*}
    \centering
    \includegraphics[width=\textwidth]{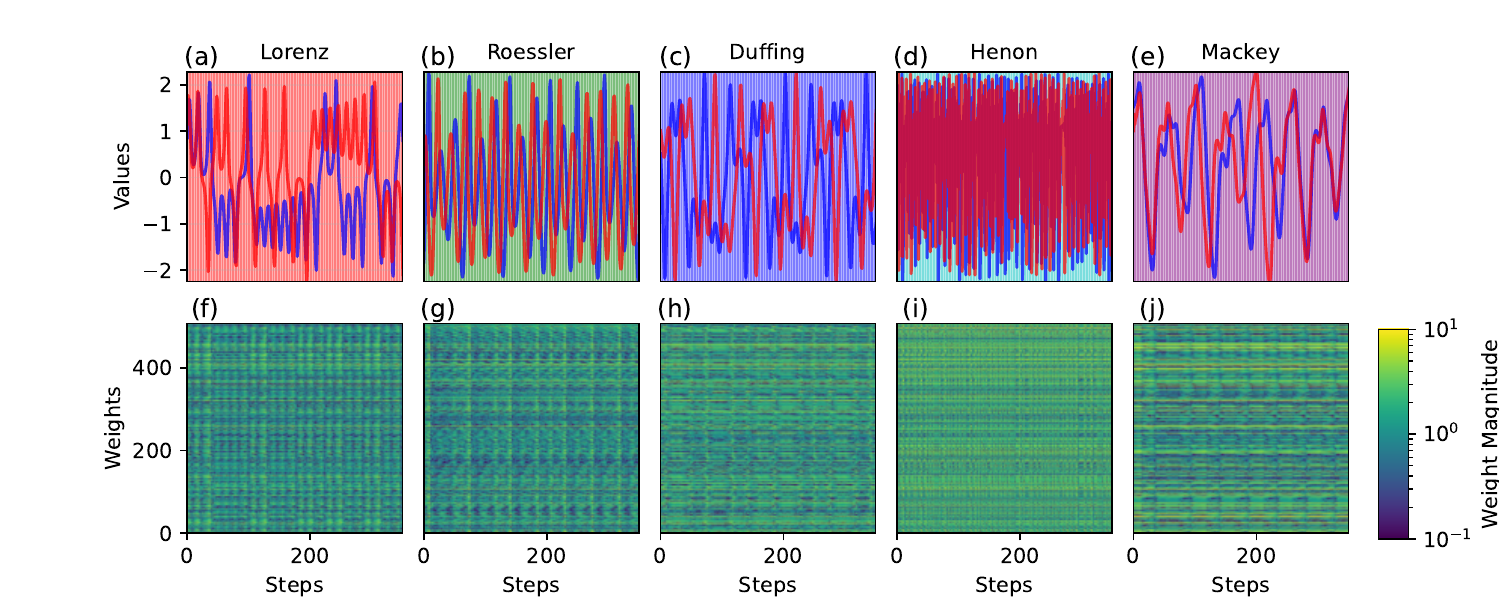}  
    \caption{Closed-loop configuration results for all five tasks using the same AERC. The upper plot shows a long time series resulting from the concatenation of predictions for all five tasks. After initializing the AERC with the starting conditions for each task, the loop was closed to allow for autonomous prediction. The lower plot represents the attention weights' magnitude plotted against the same time interval, revealing how the network dynamically allocates attention throughout the entire sequence. Different tasks trigger varying attention patterns, indicating that the network is adapting to the distinct dynamics of each system during autonomous prediction.}
    \label{fig:long_time_serie}
\end{figure*}

The top plot in Figure \ref{fig:long_time_serie} shows the results of the closed-loop configuration for all five tasks.
All five different tasks are easily reproduced, where the first steps follow the true trajectory very close, while the long term predictions seem to capture the characteristics of the five different attractors very well.

We additionally trained the AERC output layer to characterize the attractor.
The results are plotted as a colored background map in the plot, where orange, green, blue, light blue, and purple colors represent Lorenz, Rössler, Duffing, Henon, and Mackey-Glass, respectively.
It is clear, that the trained AERC manages to predict the true attractor with high precision, meaning that the AERC is able to seperate all five attractors via a representation in its embedding space, mapping the different embedded attractors via the output layer back to the original five systems.

An interesting observation is the lower plot in Fig. \ref{fig:long_time_serie}, which illustrates the attention map, representing the dynamic allocation of attention weights throughout the time series. The attention mechanism enables the network to prioritize specific nodes in the reservoir for different tasks. The attention weights vary depending on the task, showing that different sections of the network are more active for certain tasks, adapting to the distinct characteristics of each system. This adaptability reflects the versatility of the AERC in dealing with a diverse set of dynamical systems within a unified machine learning agent.

Overall, these results demonstrate that the AERC can successfully learn and predict multiple dynamical systems within the same model. The attention mechanism proves essential in allowing the reservoir to focus on task-specific dynamics, thereby enhancing the prediction accuracy across diverse chaotic systems.

We point out that this ability is strongly connected to the size of the reservoir.
For a smaller reservoir of 50 nodes, the predicted time series of the Lorenz model can be seen in  Figure \ref{fig:lorenz_time_series} of the Appendix. Here, the system has many switches between the attractors, because the embedding space is not large enough to separate all five different tasks. From the dynamical system viewpoint this is an interesting result. The embedding space of the reservoir needs to have enough volume to be able to separate the five different tasks, which are then projected down onto their three-dimensional original space. 

\subsection{Valid Prediction Time}

We show the valid prediction time (VPT) measured in Lyapunov times for each of the five tasks as a function of the reservoir size. 
The VPT measures how long the model can produce accurate predictions before diverging from the true trajectory \cite{PhysRevLett.120.024102}, defined as:
\begin{align}
    \text{VPT} = \delta_y(t) > 0.4, \quad \delta_u(t) = \frac{|y(t) - d(t)|^2}{\langle|y(t) - \langle y(t) \rangle |^2 \rangle},
\end{align}
where VPT measures the first time at which $\delta_y(t)$ surpasses the value of $0.4$, in which $\delta_y(t)$ measures the average normalized distance over the trajectory between the target $y(t)$ and the reservoirs output $d(t)$, and $\langle\rangle$ denotes the time average.

\begin{figure}
    \centering
    \includegraphics[width=0.5\textwidth]{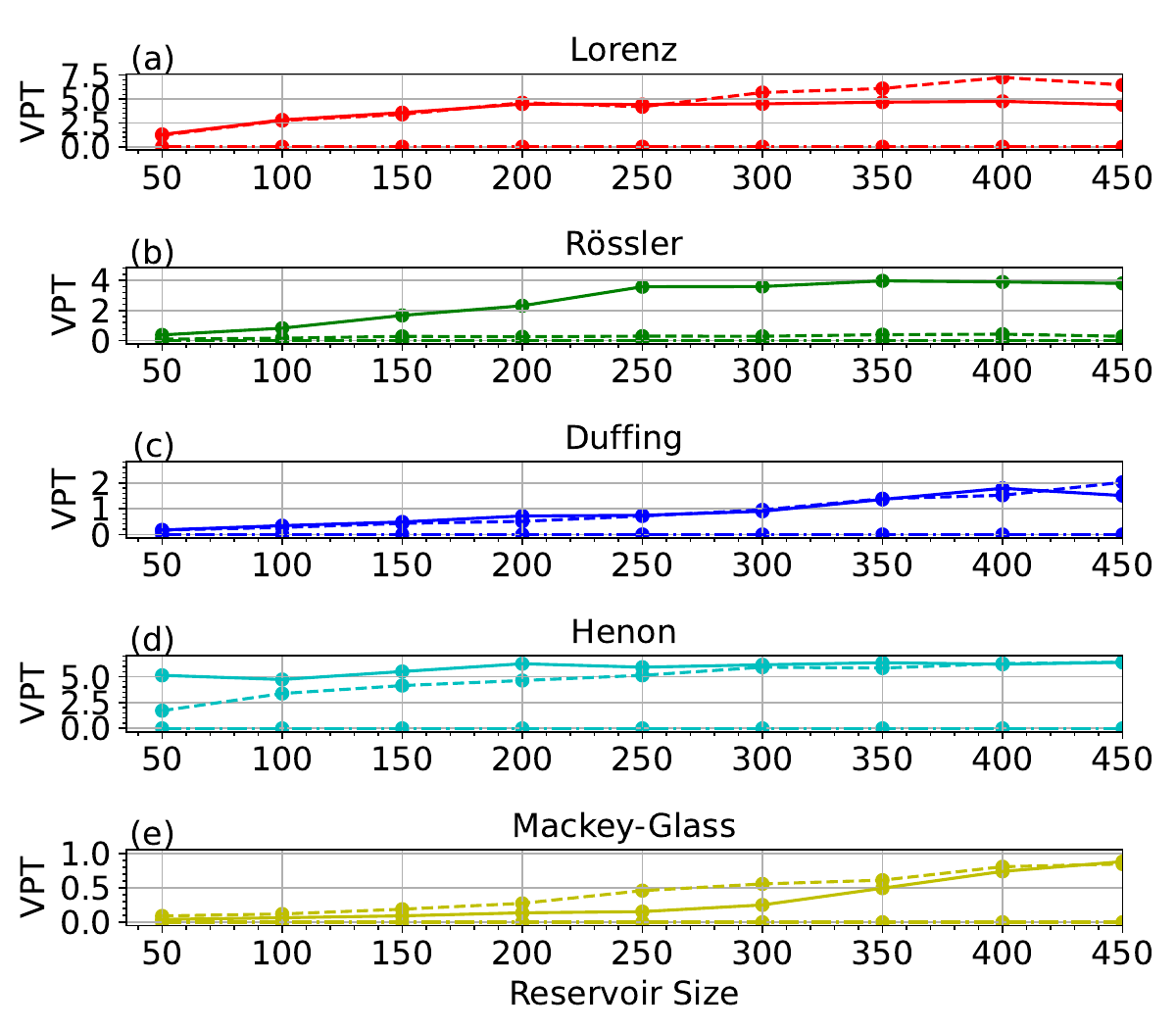}  
    \caption{Valid Prediction Time (VPT) for all five tasks as a function of the reservoir size. The solid lines depict the AERC, the dashed lines depict the classic RC approach, and the dot-dashed line depict the ridge regression model.}
    \label{fig:vpt_graph}
\end{figure}

Figure \ref{fig:vpt_graph} shows the results of VPT for the five different models. The solid curves represent the mean VPT achieved by the AERC, averaged over 50 different reservoirs.  The AERC's performance is compared against the classic ridge regression approach, where the dot-dashed line shows the ridge regression model trained on all five tasks with the same set of weights and the dashed line shows the ridge regression model trained on one single task, giving the classic approach an advantage.
As the reservoir size increases, the VPT also increase for the AERC and the single-trained classic approach.
A noteably excecption is the Rössler system where the single trained classic approach is unable to predict any significant time into the future even though it is only trained on the single task.
The ridge regression that is trained on all five tasks at the same time has a VPT of all zeros. This is not surprising, as the static weights try to average all five different tasks, which results in a chaotic mixture.
We will refrain from now on to include the ridge regression that is trained on all five tasks together, because it is a useless model.

Overall, the VPTs of the AERC and the single-trained ridge regression tend to be between 1 to 6 Lyapunov exponents. In particular, the Mackey-Glass system seems to have low prediction quality for both approaches, suggesting that the underlying reservoir architecture and chosen parameters is not optimized for the task.
Nevertheless, the question is whether an attention reservoir can learn multiple attractors at the same time, and looking at the VPT we see a performance that seems to maximize the reservoir potential.

\subsection{Spectral Similarity}

To evaluate the difference between two power spectra and histograms, we use the Pearson correlation coefficient, which measures the linear correlation between two spectra or histograms. The correlation coefficient $C$ is given by:
\[
C = \frac{\text{cov}(\mathbf{s}_1, \mathbf{s}_2)}{\sigma_{\mathbf{s}_1} \sigma_{\mathbf{s}_2}}
\]
where \( \text{cov}(\mathbf{s}_1, \mathbf{s}_2) \) is the covariance. \( \sigma_{\mathbf{s}_1} \) and \( \sigma_{\mathbf{s}_2} \) are the standard deviations of \( \mathbf{s}_1 \) and \( \mathbf{s}_2 \), respectively. A correlation coefficient close to 1 indicates a strong linear relationship between the spectra.
Examples of power spectra are shown in Fig. \ref{fig:lorenz_power_spectrum} in the Appendix.


\begin{figure}
    \centering
    \includegraphics[width=0.5\textwidth]{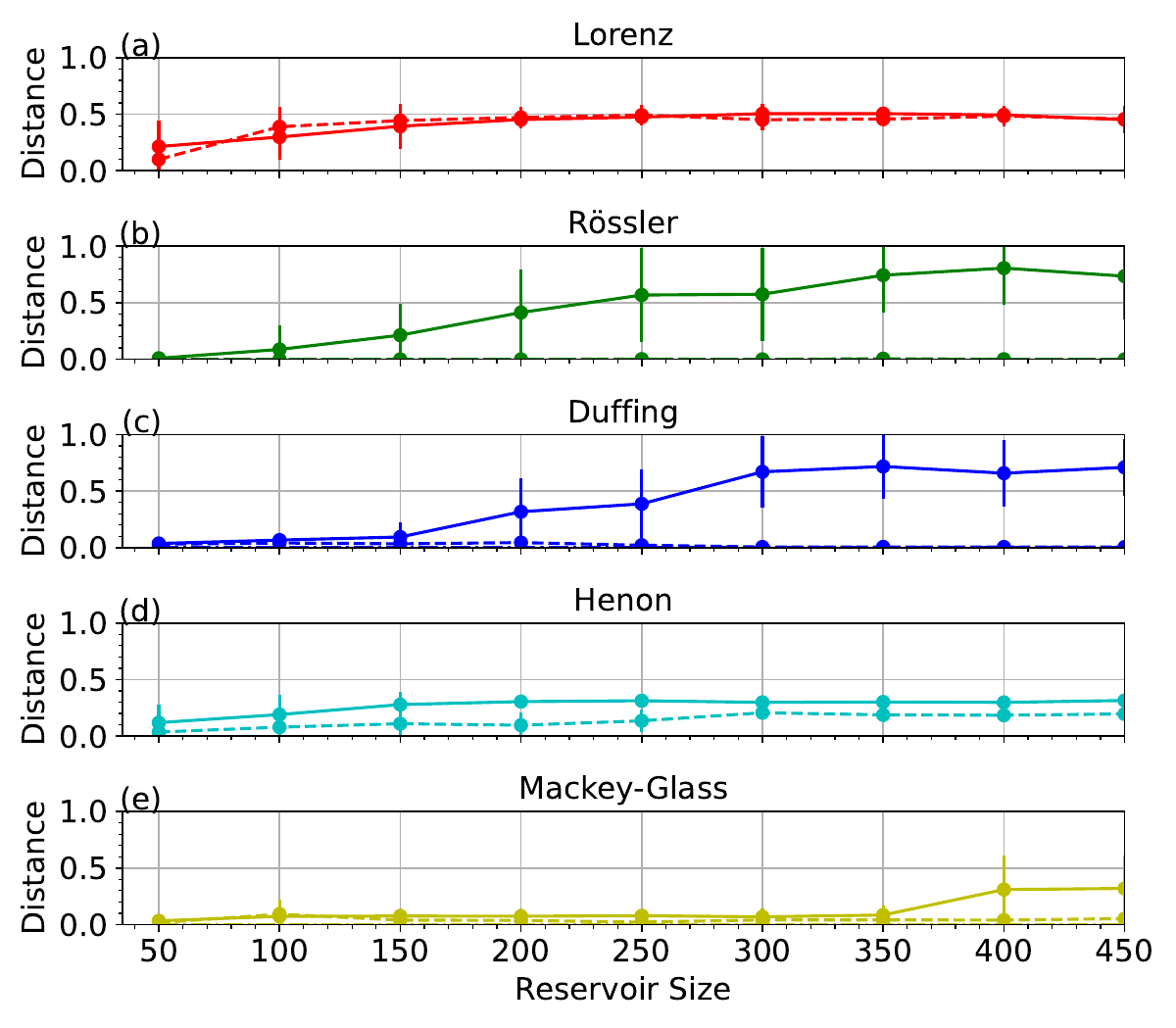}
    \caption{Spectral similarity measuered via the mean correlation coefficient over reservoir size for all five tasks. The top plot shows the AERC, while the bottom plot shows the classic RC. The colors match the system colors shown in Fig. \ref{fig:vpt_graph}.}
    \label{fig:spectrum_similarity}
\end{figure}


Figure \ref{fig:spectrum_similarity} presents the spectral similarity between the predicted and true systems as a function of reservoir size for all five tasks: Lorenz system, Rössler system, Duffing oscillator, Henon map, and Mackey-Glass delay-differential equation. The spectral similarity is evaluated using the correlation coefficient.

We show the AERC for the solid lines in Fig. \ref{fig:spectrum_similarity}. The correlation coefficient increases with reservoir size for all tasks, showing that larger reservoirs enable better reconstruction of the true system's spectral properties. It is interesting to see that systems like the Mackey-Glass seem to have to reach a certain reservoir size to show improvements. 
For the dashed lines, we show the same results for the classic approach. Comparing these results to the AERC, the AERC outperforms in mimicking the true power spectrum for all five tasks, even though the classic RC was trained only on one task at a time.

These results demonstrate that the attention mechanism allows the RC to more accurately capture the underlying dynamics of the different tasks, particularly when larger reservoirs are used. 

\subsection{Histogram Similarity}

\begin{figure}
    \centering
    \includegraphics[width=0.5\textwidth]{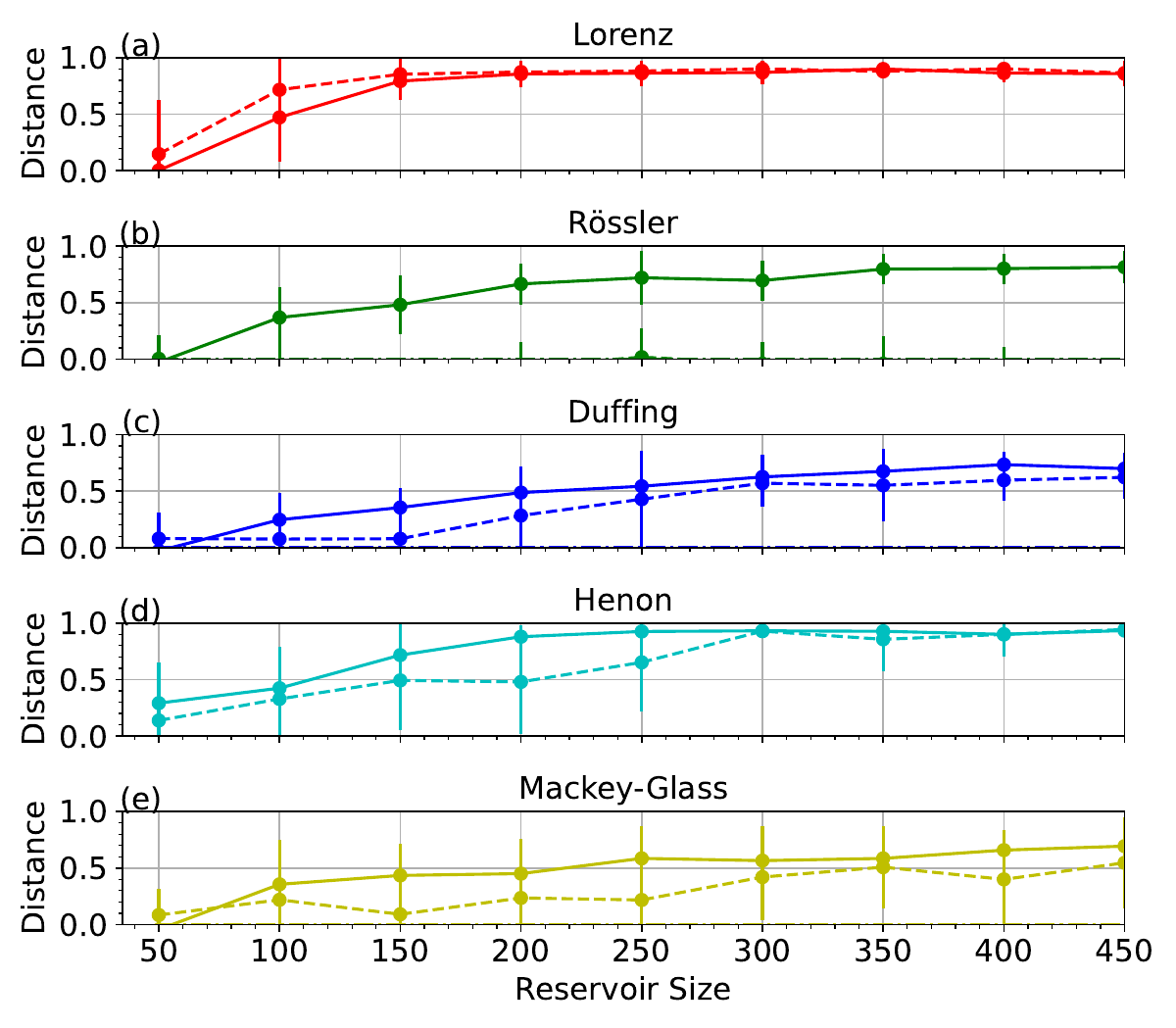}
    \caption{Same as Fig. \ref{fig:spectrum_similarity} for the histograms.}
    \label{fig:histogram_similarity}
\end{figure}

Figure \ref{fig:histogram_similarity} shows the correlation for histograms.
Examples of the histograms are shown in Fig. \ref{fig:lorenz_histograms} of the Appendix.
For the AERC the correlation coefficient increases as the reservoir size grows, indicating that the predicted state distributions become more aligned with the true distributions for larger reservoirs.
A similar graph is obtained for the single-trained ride-regression-based RC, again depicted as a dashed line.
Comparing these two graphs, the AERC slightly outperforms the ridge regression method in terms of histogram correlation, demonstrating the benefit of the attention mechanism in capturing the distributional characteristics of the different dynamical systems.
In particular, the Rössler system shows the same behavior as in Fig. \ref{fig:vpt_graph}, indicating that the classic approach has problems mimicking the true system.

These results further highlight the power of the attention mechanism in enabling the RC to better capture the distributional properties of the system states. By improving the histogram similarity between predicted and true states, the AERC demonstrates a more accurate representation of the underlying dynamics across multiple tasks.

\subsection{Discussions}

Our results demonstrate the power of the attention mechanism in RC, allowing the model to generalize across multiple tasks while maintaining or even improving the performance seen in single-task learning scenarios.
However, this comes with one caveat, as the number of trained weights drastically increases when using the AERC.
For a 500 node reservoir, we have roughly 500 times 500 weights, resulting in 250000 weights per layer. With one hidden layer that means we have 500000 weights that are trained.
We argue that this is not a limiting factor nowadays, considering that modern hardware does not have a problem running these computations.
Additionally, training can be easily performed on a modern GPU, while trained weights can be executed even on very small devices.
In our simulations, the number of weights of up to a few ten million could be trained in a few minutes on a GPU consumer (NVIDIA GeForce RTX 4080).
We imagine that a hardware-implemented reservoir computer coupled with such a neural network approach is able to perform on a wide range of tasks, while scaling better than a linear layer trained via ridge regression.

There is also much room for improvements to the scheme. For example, only a subset of reservoir states could be used for the attention layer, or the hidden layer could be reduced in size.
Because of the wide use and research on neural networks, this new hybrid approach opens doors to many known schemes from the neural network community.

\section{Controlling Multiple Attractors}
\subsection{Method}
\label{sec:switching_attractors_no_label}

By training the AERC on multiple systems an interesting property emerges. The AERC architecture presents a unique opportunity to investigate attractor switching through a control signal between attractors. We explore how the AERC performs when we use a task with transitioning from one attractor to another in a closed-loop configuration.

After training the AERC to predict the next time step for multiple systems, the model is set up in a closed-loop configuration to test its autonomous prediction capabilities. During this phase, the model's input at each time step is replaced by the predicted output from the previous time step. This closed-loop setup allows the system to operate independently of external inputs, relying solely on its internal dynamics.

To initialize the system, a short sequence of input data \( \mathbf{x}_l^a \) from one of the five dynamical systems (denoted as system \( a \)) is fed into the AERC. Once initialized, the system transitions to closed-loop operation, where the input \( \mathbf{x}_l^a \) is updated autonomously as \( \mathbf{x}_l = \mathbf{y}_{l-1} \), with \( \mathbf{y}_{l-1}\) representing the predicted output from the previous time step.

To evaluate the robustness of the AERC and its ability to adapt dynamically, a perturbation is introduced into the system during the closed-loop phase. In particular, the system input \( \mathbf{x}_l \) is altered by introducing data from another attractor (denoted as system \( b \), where \( b \neq a \)). This perturbation is applied over a control time of \( C \) steps, during which the new input becomes a weighted combination of the predicted output \( \mathbf{y}_{l-1} \) and the initial input \( \mathbf{x}_l^b \) from system \( b \) as follows:

\[
\mathbf{x}_{l} =
\begin{cases} 
(1 - \alpha) \mathbf{y}_{l-1} + \alpha \mathbf{x}_l^b, & \text{if } l \in [c_{\text{on}}, c_{\text{off}}], \\
\mathbf{y}_{l-1}, & \text{if } l \notin [c_{\text{on}}, c_{\text{off}}],
\end{cases}
\]
Here, \( \alpha \) is the control parameter to show the degree of influence between the previous attractor's dynamics and the new attractor's one. The interval $[C_{\text{on}}, C_{\text{off}}]$ is the time window of $C$ in which the control is active. When \( \alpha = 0 \), no perturbation is applied, and the system remains on the original attractor. Conversely, when \( \alpha = 1 \), the system is fully forced to the new attractor. Intermediate values of \( \alpha \) allow for a gradual transition between the two systems.


\begin{figure}
    \centering
    \includegraphics[width=0.5\textwidth]{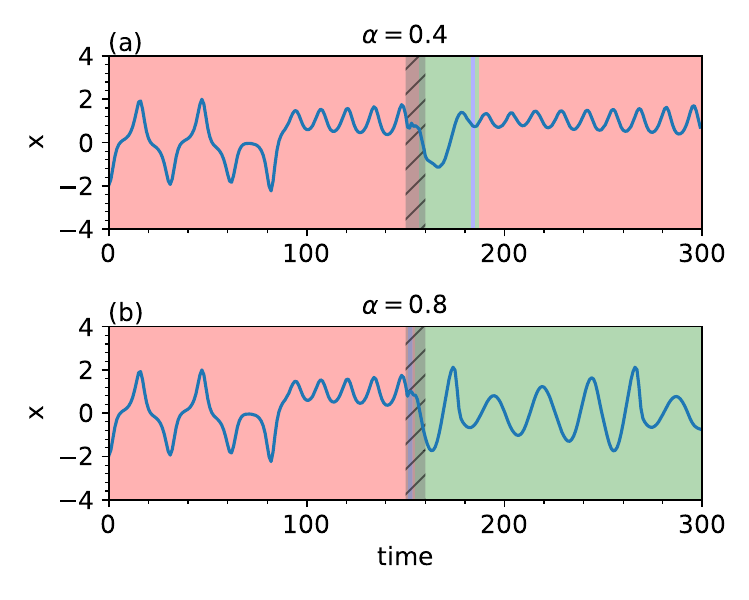}
    \caption{Time series showing the switching between the Rössler and the Duffing system for $L=12$ with (a) $\alpha=0.4$ and (b) $\alpha=0.8$. The grey area shows the control time applied.}
    \label{fig:time_series_switching}
\end{figure}

\subsection{Results of Attractor Switching}

Figure \ref{fig:time_series_switching} shows two time series of this setup with a small forcing term $\alpha=0.2$ and a large forcing term $\alpha=0.8$, respectively.
For both cases, the AERC starts in the Lorenz (red background) time series in the closed-loop configuration. At around 100, steps the control signal to transition to the Rössler system (green background) is turned on, indicated with a dashed grey area.
After switching the control signal off, the system should stay on the new attractor if the switching is successful.
For $\alpha=0.2$, a longer transition is found, where the system first is on the Duffing oscillator (blue background) after which it finally transitions to the Rössler system.
For $\alpha=0.8$, the transition is very immediate without any intermediate dynamics.


\begin{figure}[t]
    \centering
    \includegraphics[width=0.55\textwidth]{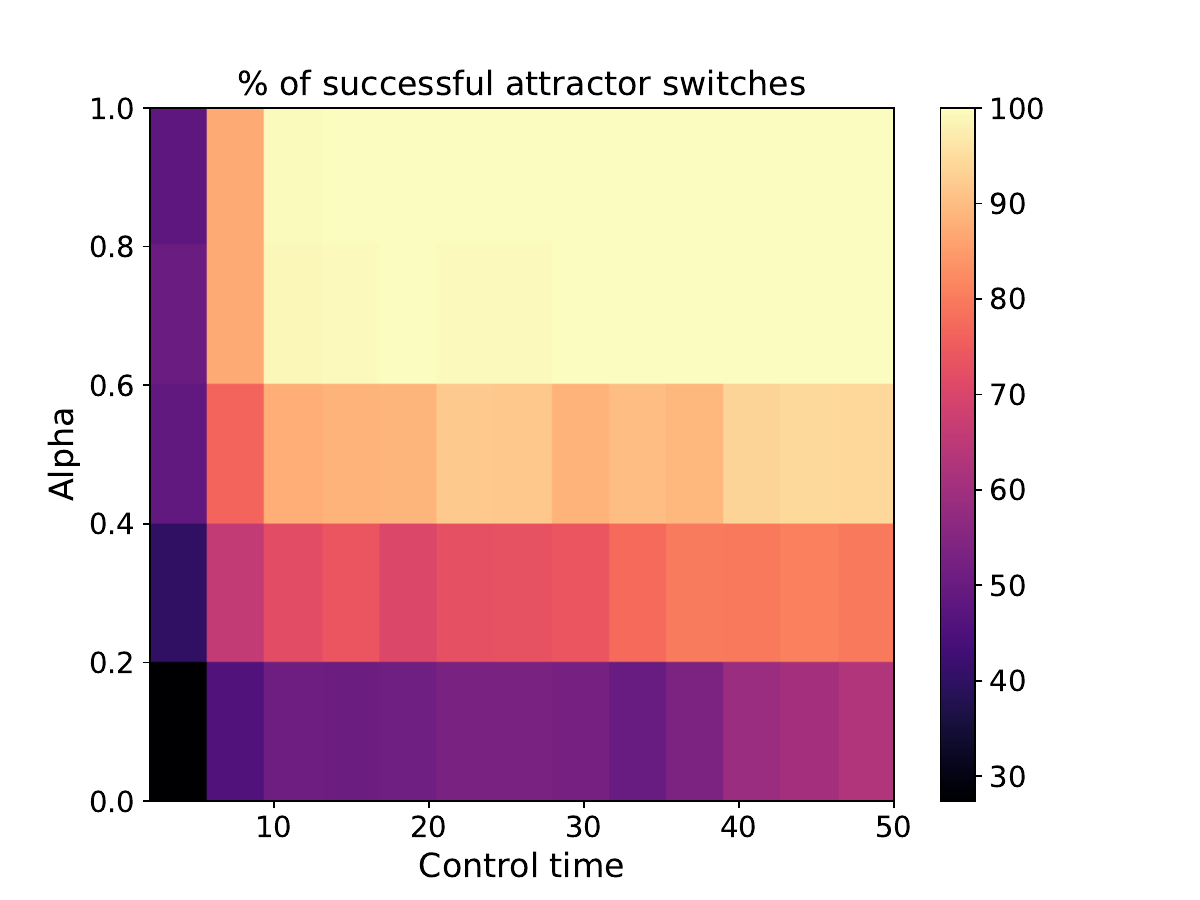}
    \caption{Successful switching rate between different attractors.}
    \label{fig:successful_switching}
\end{figure}

Next, we investigate the success rate of the switching.
We define a successful switching, when the classification of the AERC classifies the dynamics after switching to be the desired attractor. 
The results for the successful switching rate are shown in Fig. \ref{fig:successful_switching}, where the x-axis depicts the control time and the y-axis the control signal strength.
We set the reservoir size to 500 nodes.
The success rate increases with higher $\alpha$ and control times $C$.
For a reservoir size of 500 nodes, the successful switching rates can be reached up to 100\% if the control time and $\alpha$ are large enough.
The AERC manages to separate all five attractors in its embedding space, so that transition among the attractors work well if the right control time and strength are used, and spontaneous switching does not occur.

From the dynamical systems' viewpoint, this is a very interesting observation.
Considering the trained AERC as a dynamical system, this result indicates that this AERC dynamical system has (at least) five coexisting attractors, namely the Lorenz system, Rössler system, Duffing osciilator, Henon map, and Mackey-Glass equation. This system could be used as a good candidate for dynamical associative memory such as human brain \cite{Inoue2020}.

\section{Dual-Input Training Approach}
\label{sec:dual_input}

\subsection{Method}

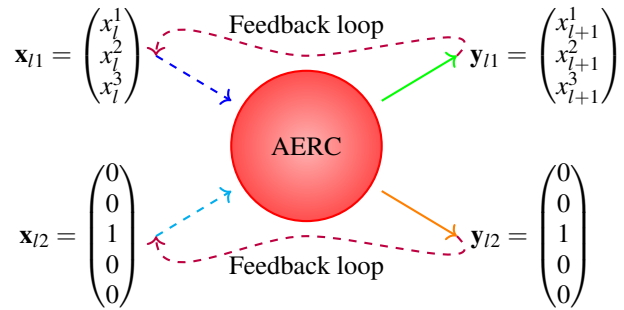
\begin{figure}
    \centering
    \begin{tikzpicture}
        \tikzstyle{input_style1} = [draw=blue, fill=blue!20, thick, dashed]
        \tikzstyle{input_style2} = [draw=cyan, fill=cyan!20, thick, dashed]
        \tikzstyle{output_style1} = [draw=green, fill=green!20, thick]
        \tikzstyle{output_style2} = [draw=orange, fill=orange!20, thick]
        \tikzstyle{circle_style} = [circle, draw=red, thick, minimum size=2cm, shading=radial, inner color=red!30, outer color=red!70]
        \tikzstyle{feedback_style} = [->, draw=purple, thick, dashed] 

        \node[circle_style] (AERC) at (0,0) {AERC};

        \draw[->, input_style1] (-2,1.2) -- (-1,0.6); 
        \draw[->, input_style2] (-2,-1.2) -- (-1,-0.6); 

        \node at (-3, 1.2) {$\mathbf{x}_{l1}= \begin{pmatrix}
x^1_l \\
x^2_l \\
x^3_l
\end{pmatrix}$};
        \node at (-3, -1.2) {$\mathbf{x}_{l2}= \begin{pmatrix}
0\\
0 \\
1 \\
0 \\
0 
\end{pmatrix}$};

        \draw[->, output_style1] (1,0.6) -- (2,1.2); 
        \draw[->, output_style2] (1,-0.6) -- (2,-1.2); 

        \node at (3.2, 1.2) {$\mathbf{y}_{l1}= \begin{pmatrix}
x^1_{l+1} \\
x^2_{l+1} \\
x^3_{l+1}
\end{pmatrix}$};
        \node at (3, -1.2) {$\mathbf{y}_{l2}= \begin{pmatrix}
0\\
0 \\
1 \\
0 \\
0 
\end{pmatrix}$};

        \draw[feedback_style] (2,1.2) to[out=45, in=0] (0,1.2) to[out=180, in=90] (-2,1.2); 
        \draw[feedback_style] (2,-1.2) to[out=-45, in=0] (0,-1.2) to[out=180, in=-90] (-2,-1.2); 
        \node[above] at (0,1.3) {Feedback loop};
        \node[above] at (0,-1.9) {Feedback loop};

    \end{tikzpicture}
    \caption{Dual-input AERC model with feedback. The input vectors $\mathbf{x}_{l1}$ and $\mathbf{x}_{l2}$ are fed into the AERC. The output consists of two vectors: $\mathbf{y}_{l1}$ for next-step prediction and $\mathbf{y}_{l2}$ for attractor classification. Feedback loops connect the upper and lower outputs back to their respective inputs.}
    \label{fig:diagram_2_AERC_dual_input_feedback}
\end{figure}

In this section, we introduce a second version of our AERC model.
We introduce a second input vector, $\mathbf{x}_{l2}$, alongside the original input vector, $\mathbf{x}_{l1}$. As shown in Fig. \ref{fig:diagram_2_AERC_dual_input_feedback}, the vector $\mathbf{x}_{l1}$ contains the dynamical system data, while $\mathbf{x}_{l2}$ represents the classes of the attractors, encoded as a one-hot vector. 
For example, the vector $(0,0,1,0,0)$ encodes the class of the Duffing oscillator.

Both input vectors, $\mathbf{x}_{l1}$ and $\mathbf{x}_{l2}$, are concatenated and then fed into the reservoir, where they are jointly processed. The new input vector to the system is now eight dimensions instead of three dimensions, where the first three dimensions are the dynamics, and the next five dimensions are the classes. 
The output consists of two vectors, $\mathbf{y}_{l1}$ and $\mathbf{y}_{l2}$. The first output vector, $\mathbf{y}_{l1}$, remains responsible for predicting the next time step of the system. The second output vector, $\mathbf{y}_{l2}$, continues to estimate the class of the attractor that the system is on, which now becomes a very trivial task for the reservoir, as it has to map the identity between the input and output vector.

By including the one-hot encoded attractor class as an additional input, the AERC is able to use this information to improve both the prediction of the next time step and the identification of the attractor. This allows for more accurate predictions, as the model has explicit information about which attractor it should be approximating, leading to improved performance in characterizing the system’s long-term behavior.

It is worth noting that this method assumes that the dataset provides a clear distinction between different tasks or attractors. However, this is not always the case, which is why we include both approaches — one with the attractor class information as input and one without it. If the attractor class information is available, it should certainly be utilized to increase information given to the machine learning agent. In cases where this information is unavailable or ambiguous, the standard approach remains valuable for learning the system’s dynamics without explicit class information.

As with the previous model, gradient descent is applied to minimize the MSE for the next-step prediction using $\mathbf{y}_{l1}$, while cross-correlation loss is used for the attractor classification using $\mathbf{y}_{l2}$.

\subsection{Results for Dual-Input Approach}

\begin{figure}
    \centering
    \includegraphics[width=0.5\textwidth]{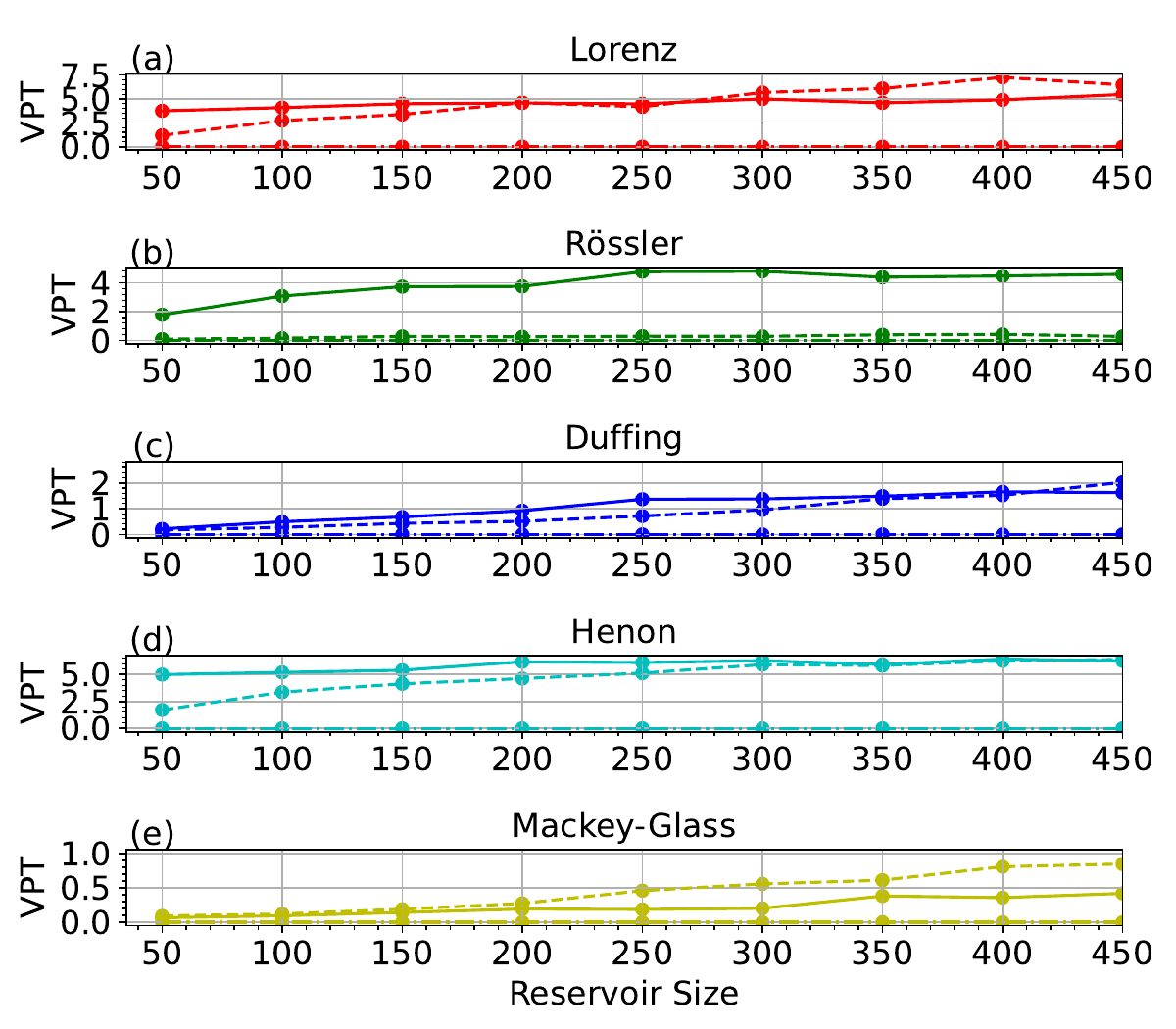}  
    \caption{Valid Prediction Time (VPT) for all five tasks as a function of the reservoir size.}
    \label{fig:vpt_graph_label}
\end{figure}

We check the VPT over the same reservoir sizes for the classic and AERC reservoir. The results are shown in Fig. \ref{fig:vpt_graph_label}.
Compared with Fig. 5, the inclusion of the class-one-hot-vector for the classic approach does not improve the prediction, as the additional information of the class does not help the classic approach at all. No new information is given to the reservoir which can be utilized to improve prediction. On the contrary, the additional input reduces the usable embedding space for the relevant information, possible also reducing the performance.
The AERC shows similar results to the approach without the dual-input. Giving the AERC the class as additional information does not seem to change short term predictions.

\begin{figure}
    \centering
    \includegraphics[width=0.5\textwidth]{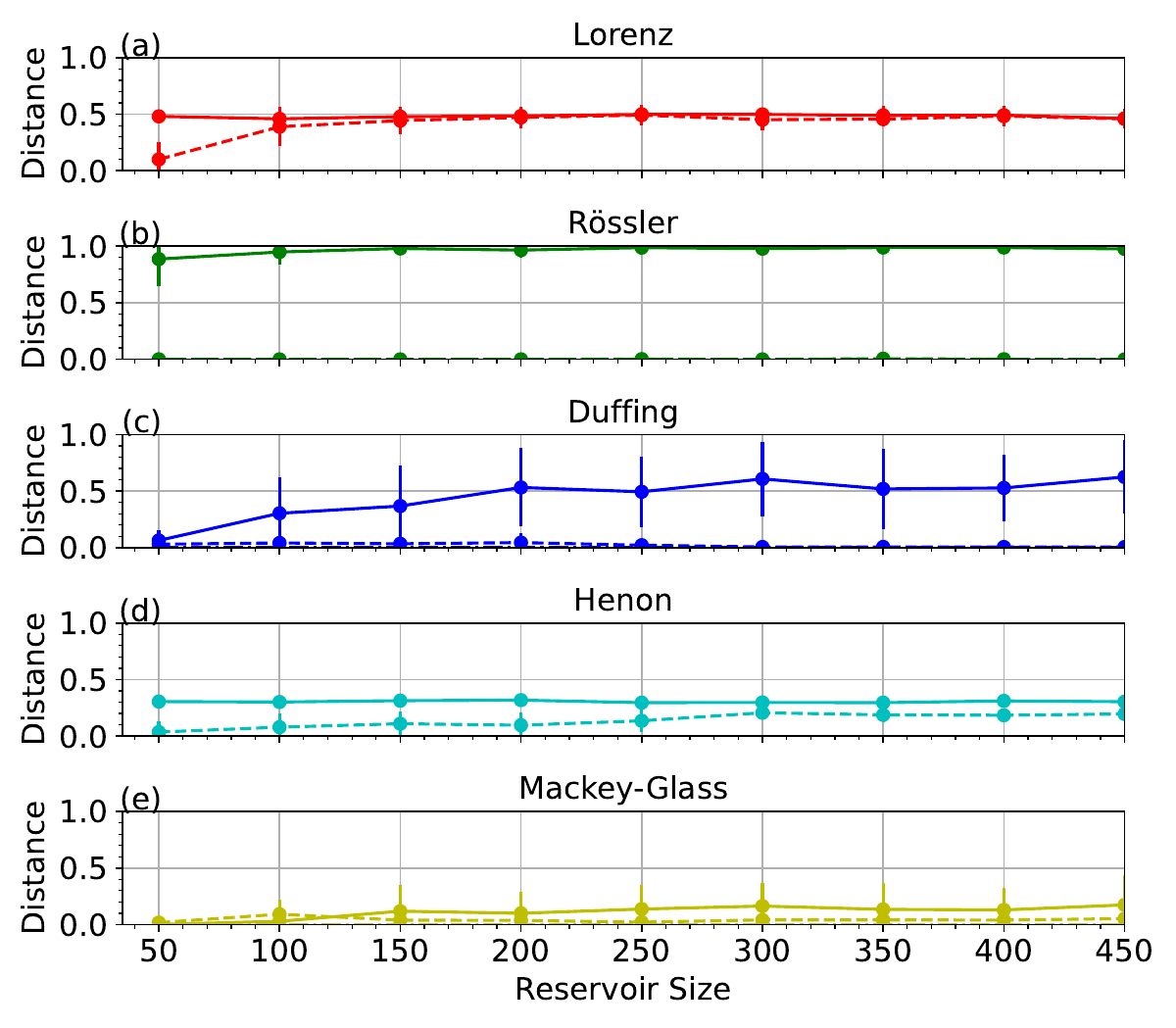}
    \caption{Same as Fig. \ref{fig:spectrum_similarity} for the spectrum with label as inputs.}
    \label{fig:spectral_similarity2}
\end{figure}

\begin{figure}
    \centering
    \includegraphics[width=0.5\textwidth]{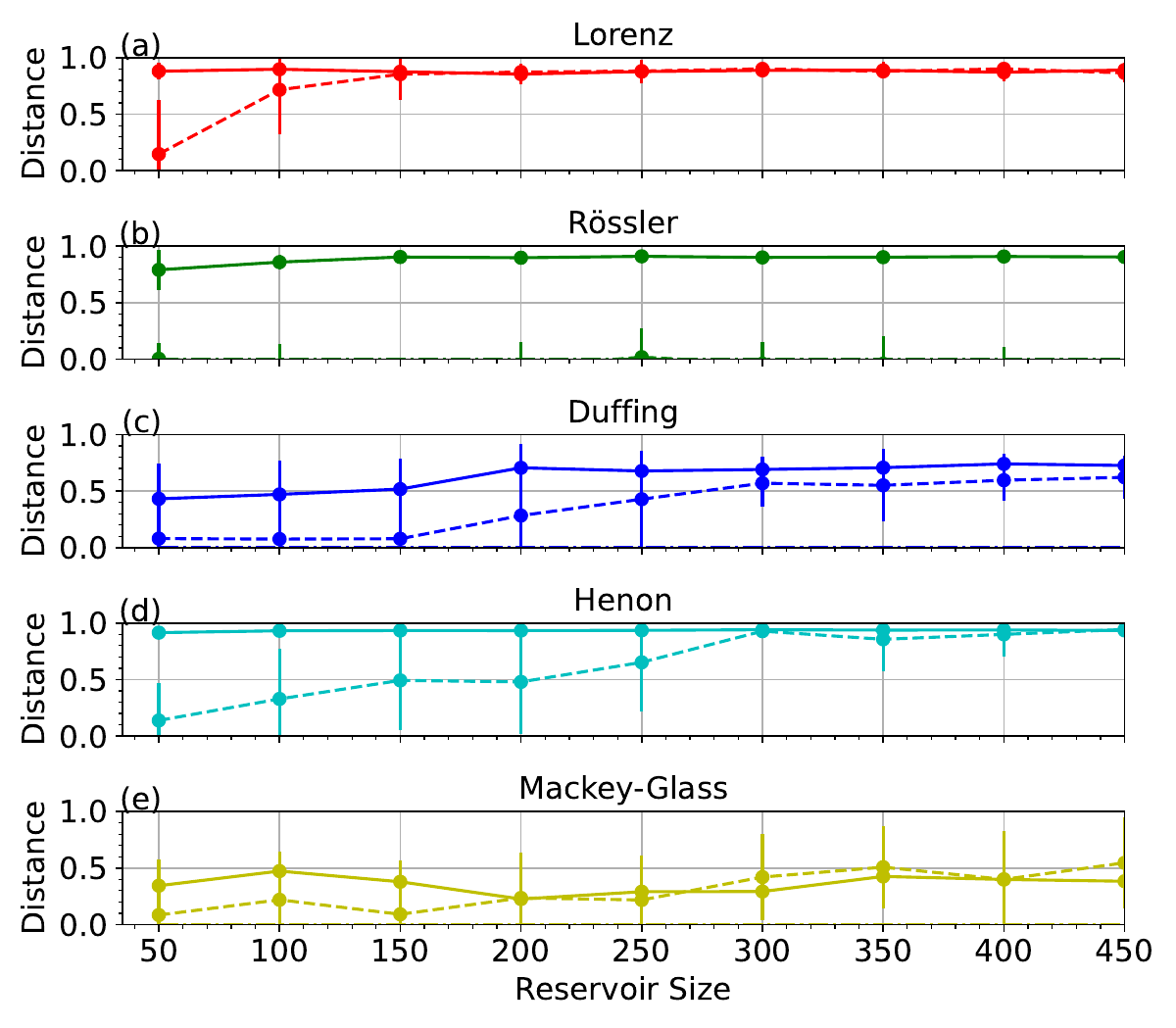}
    \caption{Same as Fig. \ref{fig:spectral_similarity2} for the histograms.}
    \label{fig:histogram_similarity2}
\end{figure}

For the long-term prediction, we show results for the spectral and histogram similarity in Fig. \ref{fig:spectral_similarity2} and Fig. \ref{fig:histogram_similarity2}, respectively.
For the classic approach, the addition of the class vector does not change the performance in simulating the long-term characteristics of the five attractors.
On the contrary, giving the AERC the additional class vector changes the performance. 
The reason is that the AERC is trained on all five tasks at the same time, thus allowing the AERC to distinguish more easily between the tasks.
Interestingly, the correlation coefficients for all five tasks seem to stay constant for this setup, showing good performance for even small reservoirs.
Surprisingly, the performance is not improved for larger reservoirs and even tends to be worse than the setup with only the dynamics input.
A reason for this could be the choice of optimization, where the gradient descent over-optimizes the classification tasks, while not giving importance to optimize the MSE for the next step prediction.
This problem could be tackled by changing the loss function in giving the MSE a higher weight.

\begin{figure}[t]
    \centering
    \includegraphics[width=0.5\textwidth]{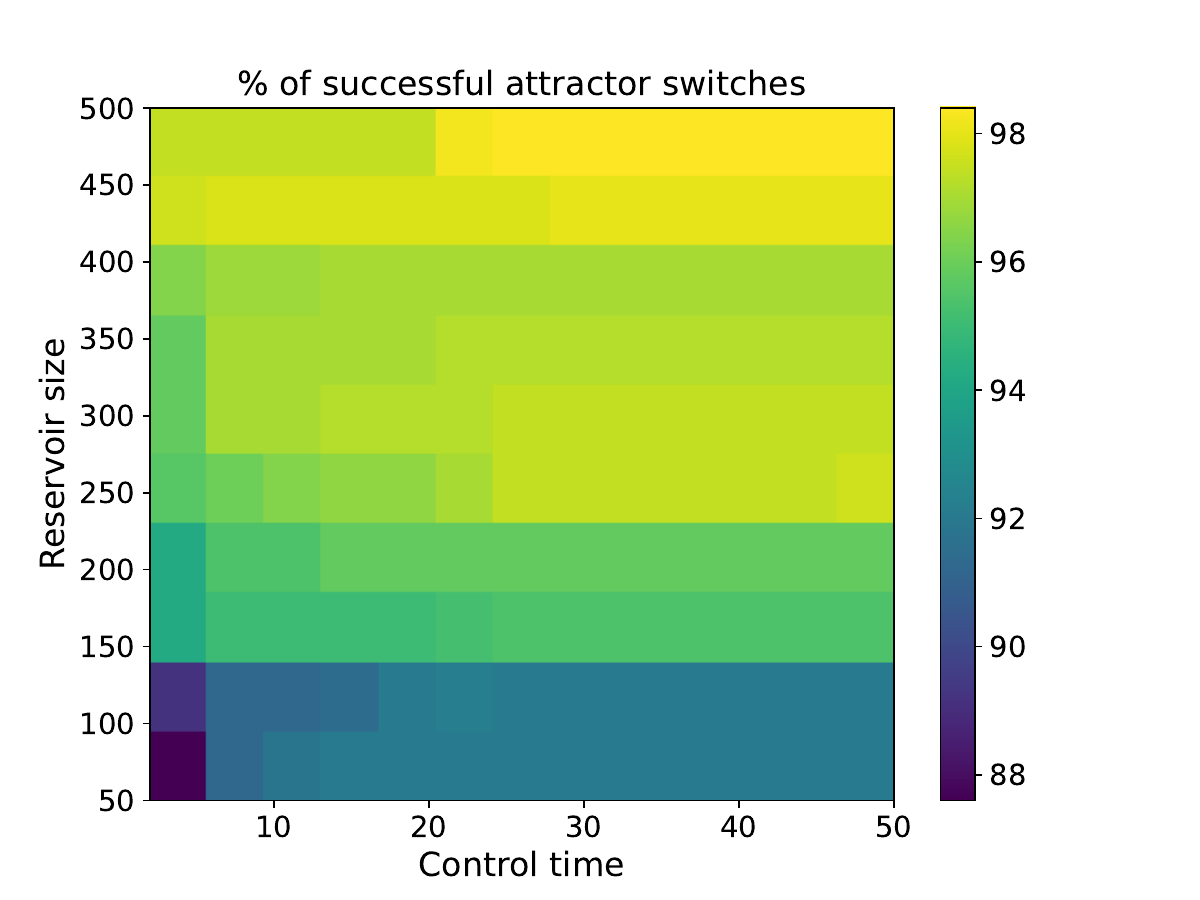}
    \caption{Successful rate of switches between attractors.}
    \label{fig:successful_switching_label}
\end{figure}

The last task will be the successful switching rate.
Like the one-input case, the system is started on one of the five attractors in the closed-loop configuration, after which a control signal is activated.
In the dual-input case though, we just switch the class of the input attractor to the desired over a few steps and then closing the loop again.
As in Fig. \ref{fig:successful_switching} the successful switches are counted and plotted in a colorcode diagram.
The results are shown in Fig. \ref{fig:successful_switching_label}, where the \textit{x}-axis shows the control time, and the \textit{y}-axis shows the reservoir size, as $\alpha$ is not used anymore due to the fact, that the one hot vector is a binary encoding of the class.
Even for small reservoirs, the success rate tends to be close to 90\%, going up to 99\% for large reservoirs and long control signals.
Contrary to the single-input case, the success rate never reaches 100\%. This is due to the fact, that systems sometimes tend to diverge with the sudden switching of attractors.
Solutions to this issue could be to introduce also an average input for the first three dimensions which encode the dynamics, or training the system directly on switching behavior.

\section{Conclusion}

In this paper we showed that the AERC can successfully predict the behavior of multiple chaotic dynamical systems simultaneously with a single set of weights. By incorporating an attention mechanism into a reservoir system, we achieved better short- and long-term prediction accuracy across various five different tasks such as the Lorenz system, Rössler system, Henon map, Duffing oscillator, and Mackey-Glass delay-differential equation.

The attention mechanism enabled the AERC to dynamically adjust its output weights based on the reservoir states, improving adaptability and enhancing prediction performance. Our results demonstrated that the AERC outperformed traditional methods, like ridge regression, in terms of VPT, spectral similarity, and histogram similarity.

Overall, we demonstrated the potential of the AERC to generalize across multiple tasks with greater accuracy than traditional models, establishing it as a powerful tool for chaotic time-series prediction. Future work could explore further applications of this architecture in other complex systems and extend its adaptability to broader tasks. There is still a lot of room to explore, tuning the hyperparameters of this hybrid system or combining it with other systems, extending its architecture.

\section{Appendix}

\subsection{Attention Layer}

Let the reservoir state for data point \( l \) be \( \mathbf{r}_l \in \mathbb{R}^N \), where \( N \) is the dimension of the reservoir state vector. This state is first transformed linearly by a weight matrix \( \mathbf{W}_1 \in \mathbb{R}^{H \times N} \), where \( H \) is the number of units in the hidden layer (which we set to \( N \) in this paper):
\[
\mathbf{z}_l = \mathbf{W}_1 \mathbf{r}_l + \mathbf{b}_1
\]
where \( \mathbf{z}_l \in \mathbb{R}^H \) represents the input to the hidden layer and \( \mathbf{b}_1 \in \mathbb{R}^H \) is a trainable bias term.

The input to the hidden layer is then passed through a ReLU activation function, which we denote as \( \sigma(\cdot) = \max(0, \cdot) \), applied element-wise:
\[
\mathbf{h}_l = \sigma(\mathbf{z}_l) = \sigma(\mathbf{W}_1 \mathbf{r}_l + \mathbf{b}_1)
\]
Here, \( \mathbf{h}_l \in \mathbb{R}^H \) is the activated hidden layer output.

The activated output \( \mathbf{h}_l \) is then linearly transformed again using a second weight matrix \( \mathbf{W}_2 \in \mathbb{R}^{N \times H} \) to produce the attention weights:
\[
\mathbf{w}_{\text{att},l} = \mathbf{W}_2 \mathbf{h}_l + \mathbf{b}_2
\]
where \( \mathbf{b}_2 \in \mathbb{R}^N \) is another bias term, and \( \mathbf{w}_{\text{att},l} \in \mathbb{R}^{N} \) provides the attention weights for one output dimension $t$ that will scale the reservoir states in producing the final output. For all output dimensions we have a seperate set of trainable weights in the attention layer.
Future discussions can be made how to optimize this procedure to reduce the number of weights.

Combining these steps, the full computation of \( \mathbf{w}_{\text{att},l} \) can be expressed as:
\[
\mathbf{w}_{\text{att},l} = \mathbf{W}_2 \sigma(\mathbf{W}_1 \mathbf{r}_l + \mathbf{b}_1) + \mathbf{b}_2
\]
$ \mathbf{W}_1$,$ \mathbf{W}_2$, $\mathbf{b}_1$, and $\mathbf{b}_2$ are combined with $\mathbf{W}_{\text{net}}$ for one output dimension.
\begin{figure*}[t]
    \centering
    \includegraphics[width=0.45\textwidth]{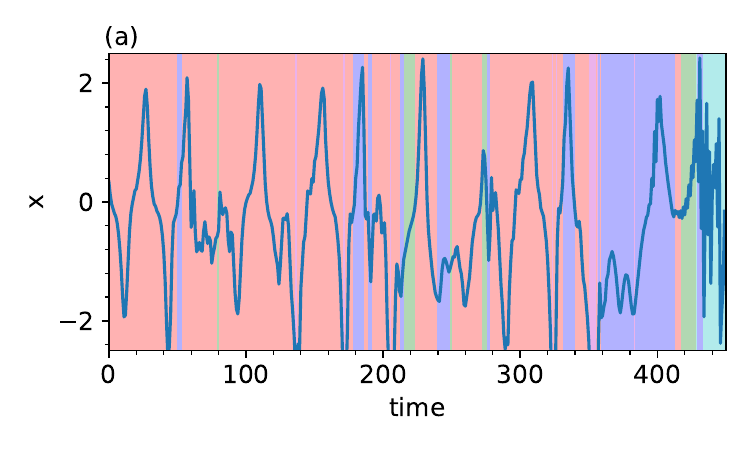}
    \includegraphics[width=0.45\textwidth]{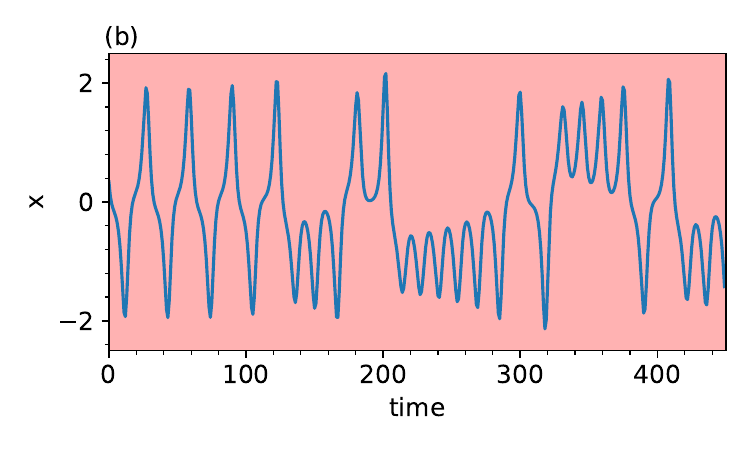}
    \caption{Time-series prediction for the Lorenz system with (a) a small reservoir (50 nodes), and (b) a larger reservoir (500 nodes). We observe that the larger reservoir size results in more accurate predictions over a longer period without any attractor switching.}
    \label{fig:lorenz_time_series}
\end{figure*}

\begin{figure*}[t]
    \centering
    \includegraphics[width=0.45\textwidth]{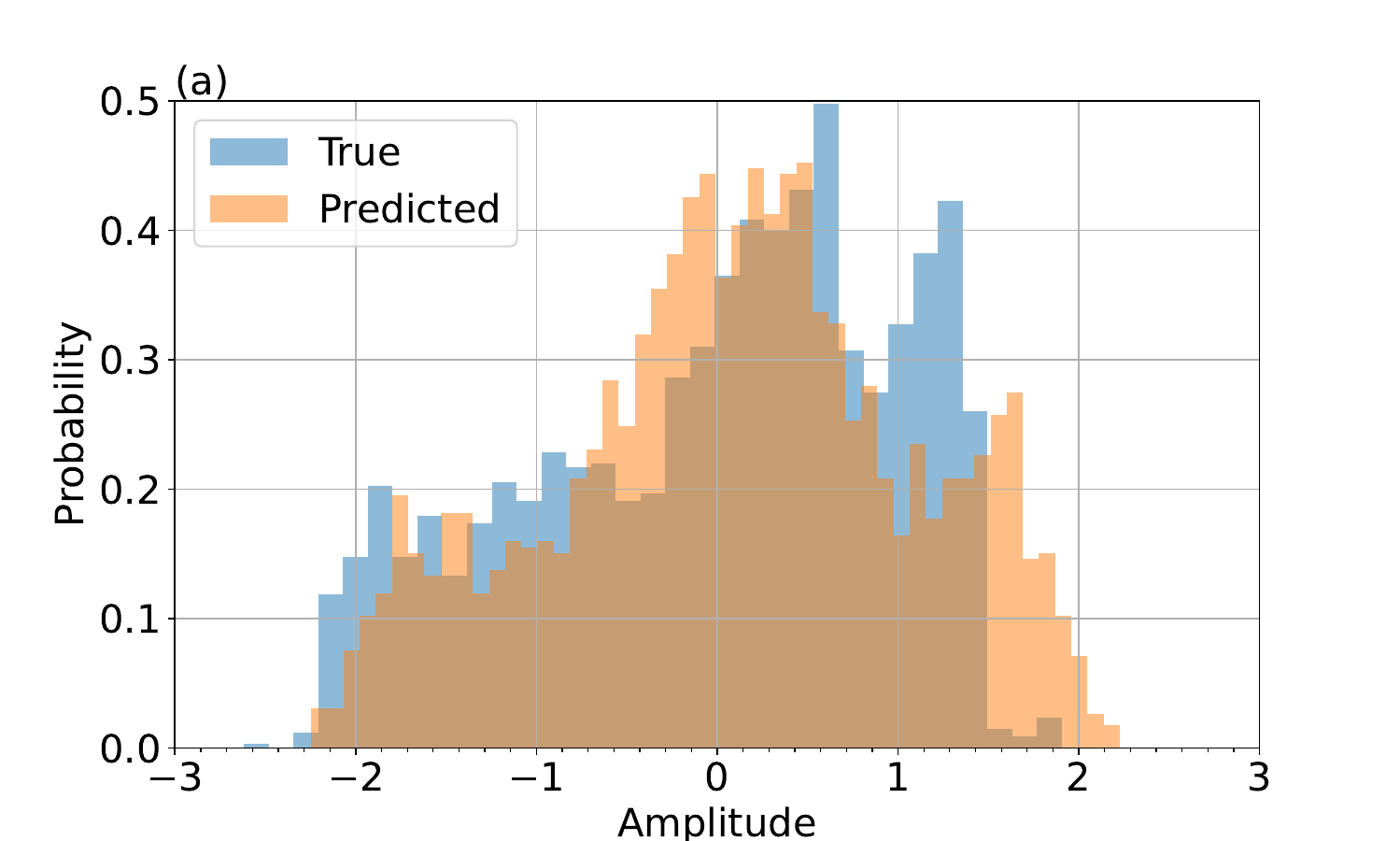}
    \includegraphics[width=0.45\textwidth]{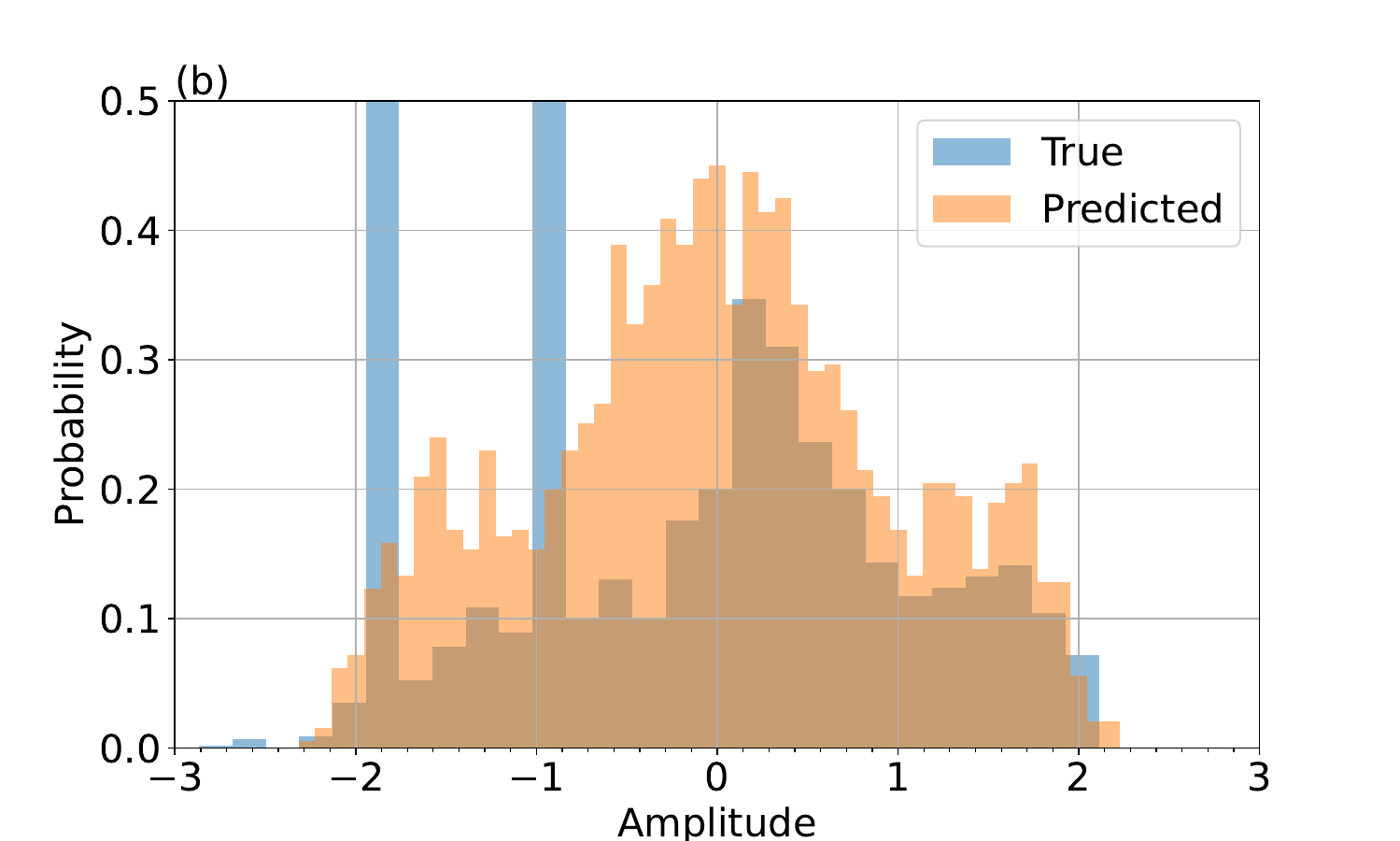}
    \includegraphics[width=0.45\textwidth]{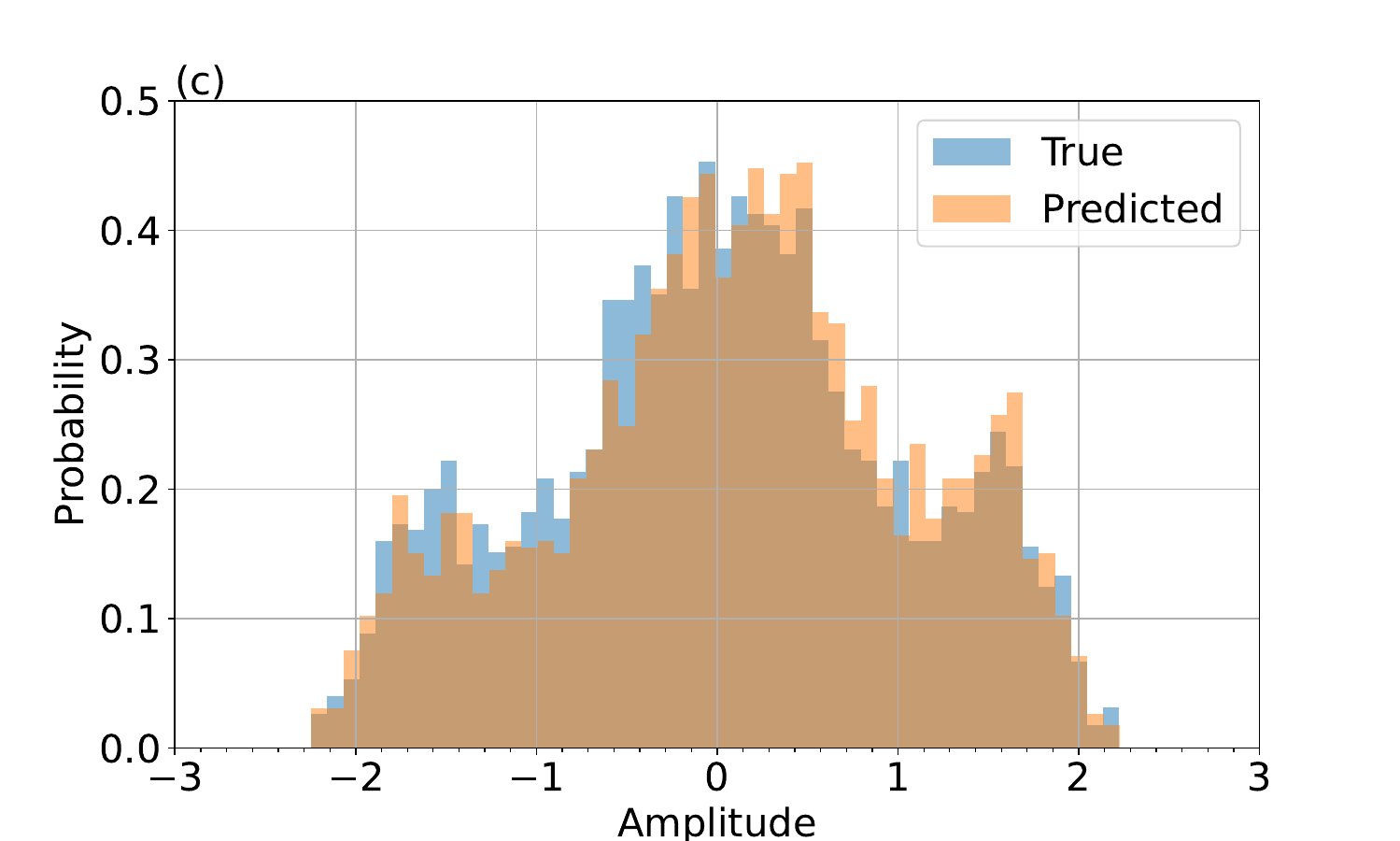}
    \includegraphics[width=0.45\textwidth]{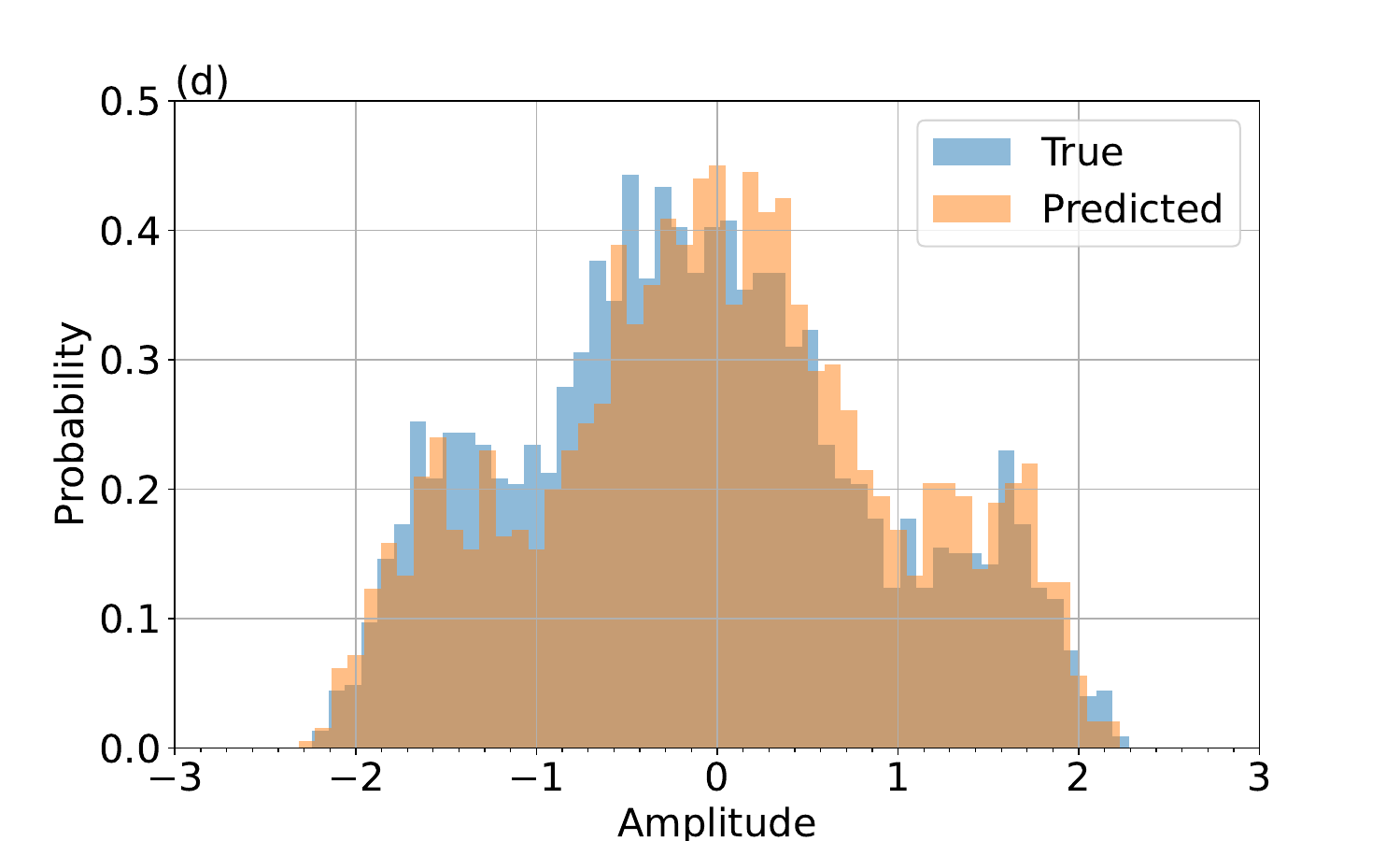}
    \caption{Histogram comparison between the true (blue) and predicted (red) signal distributions for the Lorenz system. Results for a small reservoir (50 nodes) using (a) the attention-enhanced reservoir and (b) the ridge regression approach. Results for a larger reservoir (500 nodes) using (c) the attention-enhanced reservoir and (d) the ridge regression approach.}
    \label{fig:lorenz_histograms}
\end{figure*}

\subsection{Model equations}
In this study, we simulate five dynamical systems as benchmark tasks: the Lorenz system, Rössler system, Henon map, Duffing oscillator, and Mackey-Glass delay differential equation. The detailed equations of the five models are described in the following.

\subsubsection{Lorenz System} 
The Lorenz system is a set of three differential equations originally developed to model atmospheric convection. It is known for its chaotic solutions with certain parameter values \cite{lorenz1963deterministic}: 
\begin{align*} 
\frac{dx}{dt} &= \sigma (y - x) \\ 
\frac{dy}{dt} &= x (\rho - z) - y \\ 
\frac{dz}{dt} &= x y - \beta z 
\end{align*} 
where \(\sigma\), \(\rho\), and \(\beta\) are fixed parameters. We used \(\sigma = 10\), \(\rho = 28\), and \(\beta = \frac{8}{3}\). The sampling rate is 0.05, with a total simulation time of \(375\) time units and \(7500\) sampling points.

\subsubsection{Rössler system} 
The Rössler system consists of three differential equations known for its chaotic behavior \cite{rossler}: 
\begin{align*} 
\frac{dx}{dt} &= -y - z \\ 
\frac{dy}{dt} &= x + a y \\ 
\frac{dz}{dt} &= b + z (x - c) 
\end{align*} 
with fixed parameters \(a = 0.2\), \(b = 0.2\), and \(c = 5.7\). The total simulation time is \(2000\) units, sampled at 0.27 rounded up, resulting in \(7500\) data points.

\subsubsection{Henon Map} 
The Henon map is a discrete-time dynamical system defined by the recurrence relation \cite{henon}: 
\begin{align*} 
x_{n+1} &= 1 - a x_n^2 + y_n \\ 
y_{n+1} &= b x_n 
\end{align*} 
with fixed parameters \(a = 1.4\) and \(b = 0.3\). The map is computed over \(7500\) iterations, corresponding to \(7500\) discrete time points.

\begin{figure*}[t]
    \centering
    \includegraphics[width=0.41\textwidth]{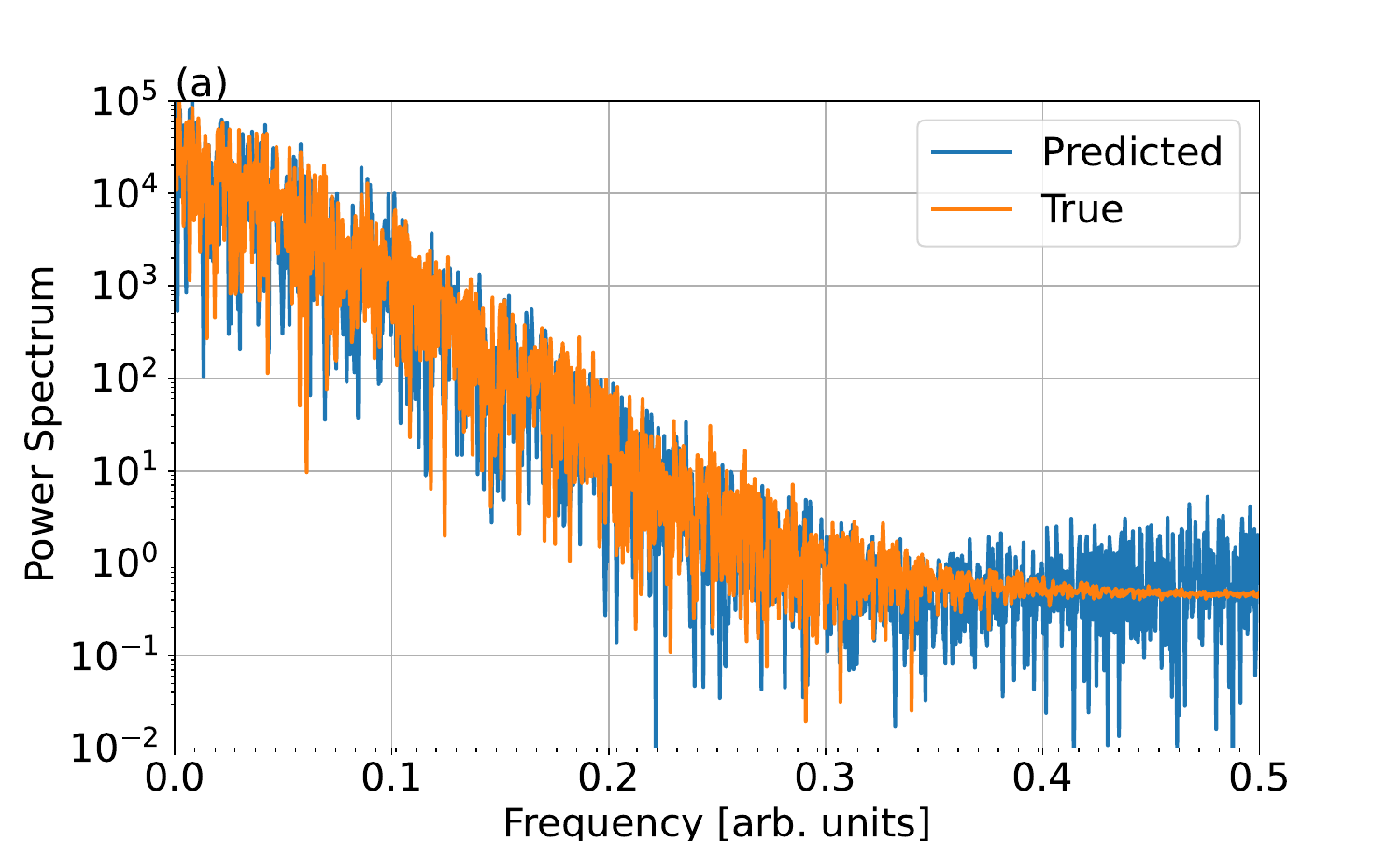}
    \includegraphics[width=0.41\textwidth]{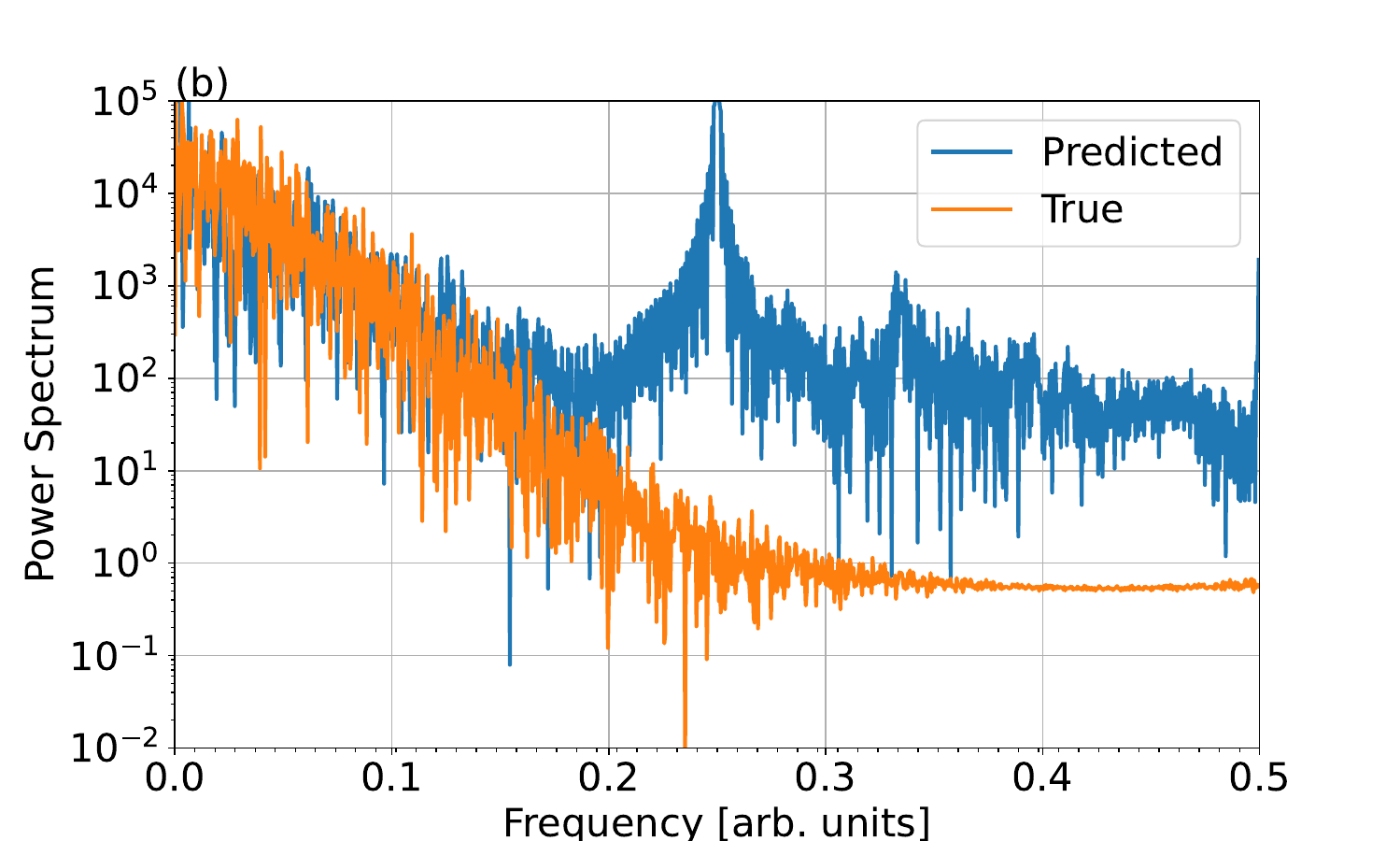}
    \includegraphics[width=0.41\textwidth]{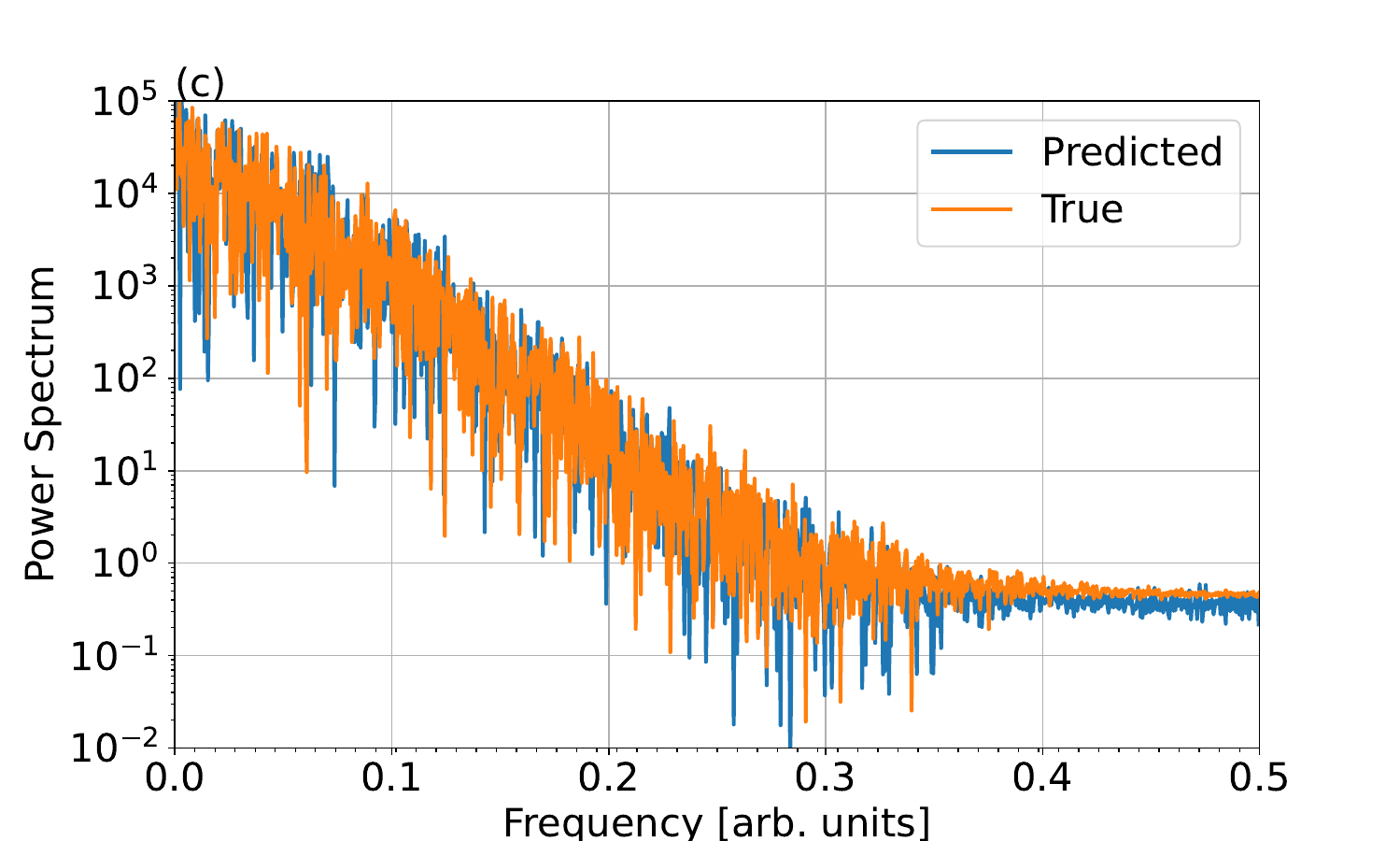}
    \includegraphics[width=0.41\textwidth]{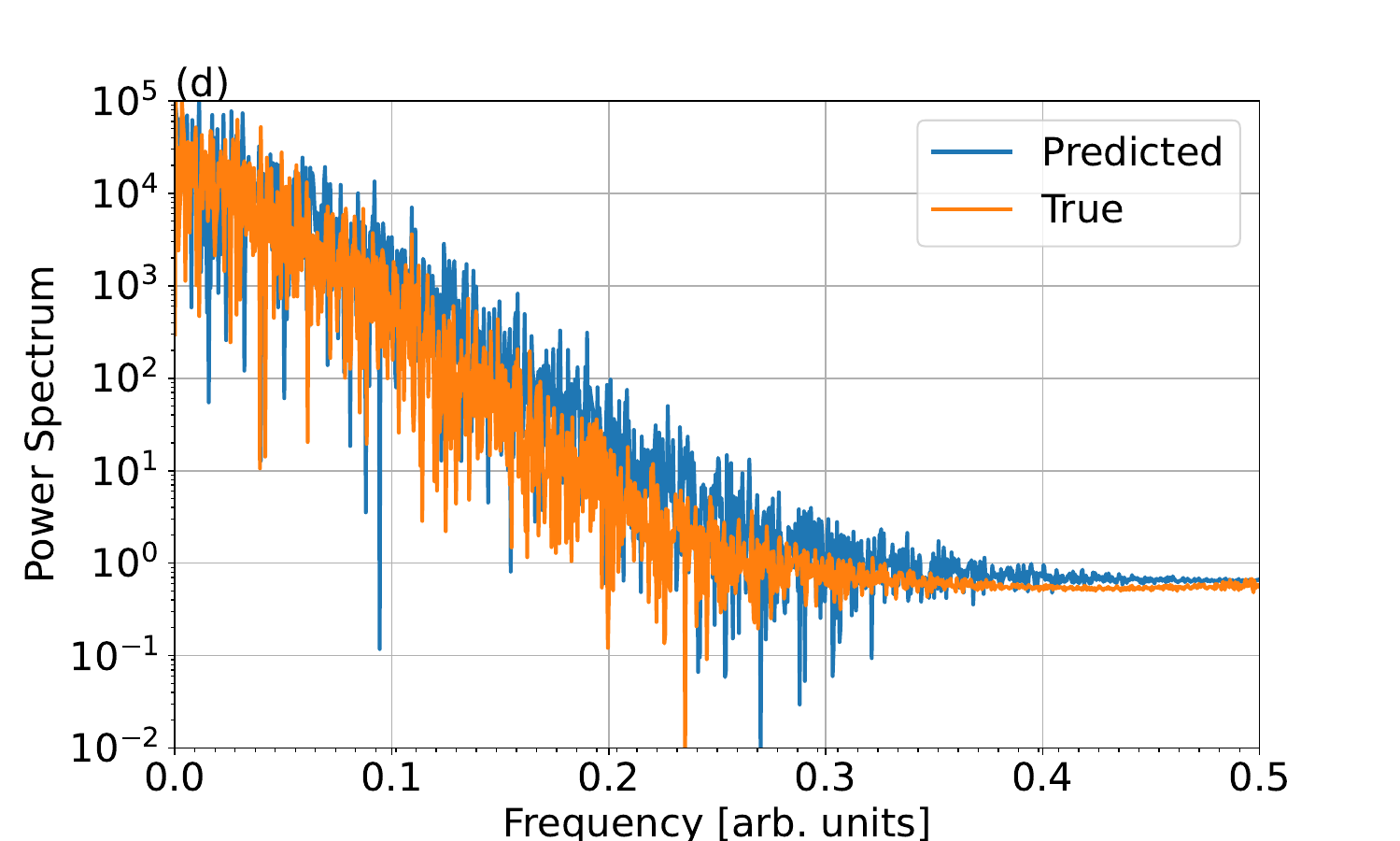}
    \caption{Power spectra comparison for the Lorenz system. Comparison for the small reservoir (50 nodes) using (a) the attention-enhanced reservoir and (b) the ridge regression approach (right). Comparison for the larger reservoir (500 nodes) using (c) the attention-enhanced reservoir and (d) the ridge regression approach.}
    \label{fig:lorenz_power_spectrum}
\end{figure*}

\subsubsection{Duffing Oscillator} 
The Duffing oscillator is a forced nonlinear second-order differential equation used to model certain damped and driven oscillators \cite{duffing}: 
\begin{align*} 
\frac{dx}{dt} &= v \\ 
\frac{dv}{dt} &= -\delta v - \alpha x - \beta x^3 + \gamma \cos(\omega t) 
\end{align*} 
with fixed parameters \(\delta = 0.5\), \(\alpha = -1.0\), \(\beta = 1.0\), \(\gamma = 0.3\), and \(\omega = 1.2\). The system is simulated over \(825\) units of time with \(7500\) data points. The sampling rate is 0.11.

\subsubsection{Mackey-Glass Delay Differential Equation} 
The Mackey-Glass equation is a delay differential equation known for its chaotic solutions \cite{mackey}: 
\begin{align*} 
\frac{dx}{dt} &= \beta \frac{x(t - \tau)}{1 + x(t - \tau)^n} - \gamma x(t) 
\end{align*} 
with fixed parameters \(\beta = 0.2\), \(\gamma = 0.1\), \(\tau = 17\), and \(n = 10\). The simulation is run over \(7500\) units of time with a total of \(7500\) data points. The sampling rate is 1.0.

All ordinary systems are integrated using Pythons Scipy library with the RK45 \cite{SciPy1.0} method and an adapted step size, while the Henon map and the Mackey-Glass delay differential equation are solved directly via the python libary ddeint \cite{ddeint}.

\subsection{Results of AERC for Lorenz System}

In the following sections, we show the results for the Lorenz system individually for a small and large reservoir size.
All AERC results are for the same set of weights for the two reservoir sizes.
The results of the ride regression approach are given by training the system for all the systems individually.
We show time-series predictions for small and large reservoirs, compare the histograms and power spectra, and display the attractor reconstruction.

\subsubsection{Time-Series Predictions}

Figure \ref{fig:lorenz_time_series}(a) and \ref{fig:lorenz_time_series}(b) show the results of time-series prediction for two reservoir sizes of 50 and 500, respectively.
The time series predictions demonstrate that the AERC improves in accuracy as the reservoir size increases. The small reservoir (50 nodes) struggles to maintain the correct trajectory, switching attractors often. As the reservoir size grows, the model stays on the attractor, leading to more precise long-term predictions.

\subsubsection{Histogram Comparisons}

Figure \ref{fig:lorenz_histograms} shows the histograms of the closed loop predictions for the 50 node reservoir (Fig. \ref{fig:lorenz_histograms}(a) and (b)) and the 500 node reservoirs (Fig. \ref{fig:lorenz_histograms}(c) and (d). Figures \ref{fig:lorenz_histograms}(a) and (c) show the AERC approach, while Figs. \ref{fig:lorenz_histograms}(b) and (d) show the ridge regression approach.
The histogram comparisons for the Lorenz system illustrate that the attention-enhanced reservoir and the ridge regression approach tend to yield distributions that are close to each other. For small reservoirs both tend to have bad matches with the true system, while the large reservoir for 500 nodes resembles the true histograms pretty closely.

\subsubsection{Power Spectrum Comparisons}

Figure \ref{fig:lorenz_power_spectrum} shows the power spectra for the 50 node reservoir (Fig. \ref{fig:lorenz_power_spectrum}(a) and (b)) and the 500 node reservoirs (Fig. \ref{fig:lorenz_power_spectrum}(c) and (d)). 
Figures \ref{fig:lorenz_power_spectrum}(a) and (c) show the AERC approach, while Figs. \ref{fig:lorenz_power_spectrum}(b) and (d) show the ridge regression approach. 
The power spectrum comparisons demonstrate that the AERC better replicates the spectral properties of the Lorenz system, particularly with larger reservoir sizes. Remember that the AERC is always trained on all five different tasks, which could explain the worse power spectra replication for the smaller reservoir compared to the Ride Regression. The increased reservoir size allows the attention-enhanced model to more accurately capture the frequency components of the Lorenz attractor.




\section{Acknowledgment}
This study was supported in part by JSPS KAKENHI (JP22H05195) and JST, CREST (JPMJCR24R2) in Japan.

\bibliography{references}

\end{document}